\documentclass[preprint]{aastex63}
\usepackage[T1]{fontenc}
\usepackage[latin9]{inputenc}
\usepackage{multirow}
\usepackage{amssymb}
\usepackage{graphicx}
\usepackage[english]{babel}

\received{16-Jan-2020}
\revised{29-May-2020}
\accepted{16-Jul-2020}

\submitjournal{AJ}

\shorttitle{Modeling chronologies and size distributions of Ceres and Vesta craters}
\shortauthors{Roig \& Nesvorn\'{y}}

\begin{document}

\title{Modeling the chronologies and size distributions of Ceres and Vesta craters}

\correspondingauthor{Fernando Roig}
\email{froig@on.br}

\author{F. Roig}
\affiliation{Observat\'{o}rio Nacional \\
Rua Gal Jos\'{e} Cristino 77, Rio de Janeiro, RJ 20921-400, Brazil}

\author{D. Nesvorn\'{y}}
\affiliation{Southwest Research Institute \\
1050 Walnut St., Boulder, CO 80302, USA}

\begin{abstract}
We infer the crater chronologies of Ceres and Vesta from a self-consistent dynamical model of asteroid impactors. The model accounts for planetary migration/instability early in the solar system history and tracks asteroid orbits over 4.56~Gy. It is calibrated on the current population of the asteroid belt. The model provides the number of asteroid impacts on different worlds at any time throughout the solar system history. We combine the results with an impactor-crater scaling relationship to determine the crater distribution of Ceres and Vesta and compare these theoretical predictions with observations. We find that: (i) The Ceres and Vesta chronologies are similar, whereas they significantly differ from the lunar chronology. Therefore, using the lunar chronology for main belt asteroids, as often done in previous publications, is incorrect. (ii) The model results match the number and size distribution of large (diameter $>90$ km) craters observed on Vesta, but overestimate the number of large craters on Ceres. This implies that large crater erasure is required for Ceres. (iii) In a model where planetary migration/instability happens early, the probability to form the Rheasilvia basin on Vesta during the last 1 Gy is 10\%, a factor of $\sim1.5$ higher than for the late instability case and $\sim2.5$ times higher than found in previous studies. Thus, while the formation of the Rheasilvia at $\sim1$ Gy ago (Ga) would be somewhat unusual, it  cannot be ruled out at more than $\simeq1.5\sigma$. In broader context, our work provides a self-consistent framework for modeling asteroid crater records.
\end{abstract}


\section{Introduction}
The crater chronology expresses the crater production rate (number of craters per unit time per surface area)
as a function of time. It encapsulates our understanding of the observed crater record on surfaces of different 
bodies. If known, it can be used to estimate the surface age, identify the dominant populations of impactors, 
and infer interesting things about the dynamical and collisional evolution of the solar system. Unfortunately,
it is quite difficult do determine an accurate crater chronology from data alone. This is because the ages of
different craters are often unknown and must be inferred by indirect means. The only crater chronology that 
is directly derived from observational data is the lunar chronology (e.g. \citealp{2001SSRv...96...55N};
\citealt{2009AJ....137.4936M}; \citealp{2014E&PSL.403..188R}). The Moon has a well preserved crater record 
and the soil samples returned by lunar missions can be used to infer accurate absolute ages of at least some 
lunar craters and basins. This provides time anchors from which the lunar chronology can be reconstructed. 

For most other solar system bodies, for which the crater record is not well preserved and/or 
the absolute crater ages are unknown, the crater chronology must be inferred by different means (e.g. \citealp{2010P&SS...58.1116M};
\citealp{2012P&SS...66...87M}; \citealp{2014P&SS..103..131O}; \citealp{2016NatCo...712257M}). For example, 
some researchers have re-scaled the lunar chronology to other bodies (\citealp{2009AJ....137.4936M};
\citealp{2014P&SS..103..104S}), including the main belt asteroids, even if this method may be difficult to 
justify (\citealp{2014P&SS..103..131O}). Another approach, which we pursue here, is to model the evolution
of impactors and their impacts on target bodies, and use the scaling laws (\citealp{2007Icar..187..345H}; 
\citealp{2016JGRE..121.1695M}; \citealp{2016Icar..271..350J}) to determine the expected crater distributions.
The results are then compared to observations.

Before 2011 our knowledge of the asteroid crater records was based on spacecraft images of $\sim$10 of these bodies, 
all of them smaller than 100 km in diameter. The arrival of the \textit{Dawn} spacecraft to Vesta in 2011 and 
Ceres in 2015 opened a new window into studies of impact cratering in the asteroid belt. A large basin 
on Vesta's surface have been suggested to explain the Vesta's collisional family \citep{1993Sci...260..186B}. It was later 
imaged by the Hubble Space Telescope \citep{1997Icar..128...88T} and \textit{Dawn}, and found to be $\simeq 500$
km in diameter (named Rheasilvia). \textit{Dawn} has also discovered another basin on Vesta, now called Veneneia, 
roughly 400 km in diameter \citep{2012Sci...336..690M}. In contrast, Ceres's surface does not show any obvious basins 
and the largest craters, Kerwan and Yalode, have diameters of only 280 km and 260 km, respectively 
\citep{2016NatCo...712257M}.
This is puzzling because Ceres has a collisional cross-section that is $\sim 4$ times larger than Vesta.
For two of Vesta's basins, there should thus be $\sim 8$ basins on Ceres, but there is none.

Previous attempts to derive a crater chronology for Vesta have been carried out by \citet{2014P&SS..103..104S} 
and \citet{2014P&SS..103..131O}. The former work used the lunar chronology and rescaled it, by simply multiplying
the crater production rate by a fixed factor, to Vesta. They estimated the Rheasilvia and Veneneia age to be 
$\sim3.5$ Gy. This is a significantly older age of Rheasilvia than the one ($\sim1$ Gy) suggested in 
\citet{2012Sci...336..690M}. At least part of this difference is due to different crater counting strategies
adopted by different research teams.
The young age of Rheasilvia would be more in line with the age of the Vesta family,
thought to form in aftermath of the Rheasilvia impact, which was estimated from arguments based on the collisional 
grinding of family members \citep{1996A&A...316..248M}. Dynamical modeling of the Vesta family does not constrain 
the family age well and admits ages $\geq1$ Gy (\citealt{2005A&A...441..819C}; \citealp{2008Icar..193...85N}),
which are compatible with either age estimate mentioned above. 

\citet{2014P&SS..103..131O} developed a new chronology for Vesta based on a synthesis of previous results. Their
chronology accounts for the long-term dynamical depletion of the asteroid belt \citep{2010Icar..207..744M}, 
effects of planetary migration/instability and scattering by planetary embryos that may have resided in 
the belt during the earliest stages (\citealp{2001Icar..153..338P}; \citealp{2007Icar..191..434O};
\citealp{2010AJ....140.1391M}). 
Their chronology implies the Rheasilvia age to be $\sim1$ Gy and
creates some tension with the low probability of forming Rheasilvia this late ($\sim4$\% according 
to \citealp{2014P&SS..103..131O}). They also pointed out a significant difference between the lunar and Vesta chronologies, 
suggesting that the flux of impactors on Vesta was {\it not} orders of magnitude higher during the lunar 
Late Heavy Bombardment (LHB). 

A similar analysis was published for Ceres in \citet{2016Sci...353.4759H} and \citet{2016NatCo...712257M}. The 
former work applied both the lunar and O'Brien chronologies to Ceres and determined a relatively young age
of the Kerwan crater (550-720 My). The absence of (obvious) large basins on Ceres is puzzling. \citet{2016NatCo...712257M} 
proposed that some large depressions observed on the Ceres's surface, referred to as \textit{planitia}, could be strongly relaxed basins. They identified at least two of these topological features, Vendimia planitia with
a $\sim830$ km diameter and another planitia with a $\sim570$ km diameter. Various geological mechanisms related to crustal relaxation, including potentially recent geologic activity, could be responsible for nearly complete basin erasure. 

Here we determine the crater chronologies of Ceres, Vesta and the Moon using a dynamical model of the asteroid belt 
from \citet{2017AJ....153..103N}. See that work for a complete description of the model. In brief, 
the model accounts for the early dynamical evolution of the belt due to migration/instability of the outer planets
and tracks asteroid orbits to the present epoch. The main asteroid belt, well characterized by modern surveys,   
is then used to calibrate the number and orbits of asteroids at any given time throughout the solar system history. 
The model does not account for other effects, such as scattering by planetary
embryos, or other impactor populations, such as comets, leftovers of the terrestrial planet accretion, etc. 
In Sect. \ref{sec:The-model}, we describe the model in more detail and explain the method that we used to 
determine the crater chronology and size distribution. The results for Vesta and Ceres are discussed in 
Sect. \ref{sec:Results}. Sect. \ref{sec:Conclusions} summarizes our main conclusions.

\section{Model\label{sec:The-model}}
\subsection{Dynamical model\label{sec:Dynamical-model}}
We use the dynamical model of \citet{2017AJ....153..103N} to determine the crater chronologies of Ceres and Vesta.
In that work, we performed a numerical simulation --labeled as CASE1B-- of 50,000 test asteroids over the age of 
the solar system. The simulation starts at the time of the solar nebula dispersal (it does not account for gas drag).
The adopted physical model takes into account gravitational perturbations of all planets from Venus to Neptune (Mercury is 
included for $t \leq t_{\mathrm{inst}}$, where $t_{\mathrm{inst}}$ is the time of dynamical instability; see below).
During the early stages, the giant planets are assumed to evolve by planetesimal-driven migration and dynamical 
instability (the so-called jumping-Jupiter model; \citealp{2009A&A...507.1041M}; \citealp{2012Natur.485...78B}; 
\citealp{2012AJ....144..117N}). See \citet{2018ARA&A..56..137N} for a review. The simulations span 4.56 Gy and the time of 
the instability time $t_{\mathrm{inst}}$ is considered to be a free parameter. The Yarkovsky effect and collisional 
evolution of the main belt is not modeled in \citet{2017AJ....153..103N}. This limits the reliability of the model
to large asteroids for which these effects are not overly significant \citep{2018AJ....155...42N}. Comets and other 
impactor populations are not considered. This is equivalent to assuming that Ceres and Vesta crater records are
dominated by asteroid impactors. 

The dynamical model of \citet{2017AJ....153..103N} employed a flexible scheme to test any initial orbital distribution
of asteroids. By propagating this distribution to the present time and comparing it with the observed distribution
of main belt asteroids, we were able to reject models with too little or too much initial excitation (also see 
\citealp{2015AJ....150..186R}). From the models that passed this test we select the one that has the Gaussian distributions in 
$e$ and $i$ with $\sigma_e=0.1$ and $\sigma_i=10^\circ$, and a power-law radial surface density $\Sigma(a)=1/a$. 
We also tested other initial distributions, such as the one produced by the Grand Tack model \citep{2011Natur.475..206W},
and will briefly comment on them in Sect. 3. The Grand Tack distribution is wider in eccentricity (approximately 
Gaussian with $\sigma_e \simeq 0.2$ and Rayleigh in $i$ with $\sigma_i \simeq 10^\circ$; see \citealp{2015AJ....150..186R} for explicit definitions of these distributions). 

The impact probability and velocity of a test asteroid on a target world is computed by the \"Opik algorithm 
\citep{1994Icar..107..255B}. This allows us to account for impacts on bodies that were not explicitly included in the 
simulation, such as Ceres, Vesta or the Moon. Ceres and Vesta are placed on their current orbits since time zero
(corresponding to the dispersal of the protosolar nebula). This is only an approximation because in reality both
these asteroids must have experienced orbital changes during the planetary migration/instability. Establishing how
these changes may have affected their crater records is left for future work. See \citet{2017AJ....153..103N} for the 
method used for the Moon. The impact probabilities are initially normalized \textit{to 1 test particle surviving 
at the end of the simulation}. In other words, the impact probabilities directly provided by a given simulation are divided by the total number of test particles that survived at the end of that simulation. This normalization is necessary, because the final state of the simulation resembles well the present asteroid belt only in terms of orbital distribution, but not in absolute numbers. The actual impact flux is obtained by multiplying these normalized impact probabilities by the number of asteroids larger than a given size in the present asteroid belt (see Eq. (\ref{eq:ntd}) below).

\subsection{Crater chronology\label{subsec:chrono}}
The usual approach to modeling crater records of planetary and minor bodies consists of two steps. In the first step, 
scientists define the chronology function, $f(t)$, which gives the crater production rate as a function of 
time $t$. In the second step, the model production function (MPF), $n(D_{\mathrm{crat}})$, is synthesized from available 
constraints to compute the crater production rate as a function of crater diameter, $D_{\mathrm{crat}}$. The number of craters 
is then computed as $n(t,D_{\mathrm{crat}})=f(t)\,n(D_{\mathrm{crat}})\,{\rm d}t\,{\rm d}D_{\mathrm{crat}}$. Integrating this relationship over $t$ and/or  
$D_{\mathrm{crat}}$ leads to cumulative distributions (e.g., the number of craters larger than diameter $D_{\mathrm{crat}}$ produced since time $t$). 
This approach implicitly assumes that MPF is unchanging with time which may not be accurate if size-dependent 
processes such as the Yarkovsky effect \citep{2015aste.book..509V} influence the impactor population. We do not
investigate such processes here.

Here we use a notation where $t$ measures time from time zero, corresponding to the dispersal of the protosolar gas 
nebula, to the present epoch ($t=0$ to 4.56 Gy) and $T$ measures 
time backward from the present epoch to time zero; thus $T=4.56\,{\rm Gy}-t$. We first define the chronology function and MPF 
in terms of the {\it impactor} flux and diameters (the conversion method from impactor properties to craters is
described in Sect. \ref{subsec:mpf}). The cumulative number of impacts, $n(T,D_{\mathrm{ast}})$, of asteroids larger than the diameter $D_{\mathrm{ast}}$ in the 
last $T$, is
\begin{equation}
n(T,D_{\mathrm{ast}})=F(T) \, \mathcal{N}(>\!\! D_{\mathrm{ast}})\label{eq:ntd}
\end{equation}
where $\mathcal{N}(>\!\!\! D_{\mathrm{ast}})$ is the current number of main belt asteroid larger than $D_{\mathrm{ast}}$ and $F(T)$ is the cumulative 
chronology function obtained from the dynamical model (here normalized to one asteroid larger than $D_{\mathrm{ast}}$ at $T=0$). 
Eq. (\ref{eq:ntd}) represents a forward-modeling approach that is independent of any crater data; instead, it relies 
on the accuracy of numerical simulations to reproduce the main belt evolution and our understanding of the
main belt size distribution (see Sect. \ref{subsec:mpf}). 

Having the chronology function, the intrinsic impact probability (actually, the expected value of a Poisson distribution) on the target world, $P_{i}$, can be obtained as:
\begin{equation}
P_{i}(T)=\frac{4\pi}{S}\frac{{\rm d}F(T)}{{\rm d}T}\label{eq:pit}
\end{equation}
where $S$ is the surface area of the target and the factor $4\pi$ accounts for the difference between the surface area 
and the cross section. With this definition of $P_{i}$, the total number of impacts is given as $P_{i}\,R^{2}\,n\, \Delta t$, 
where $R$ is the target radius, $n$ is the number of impactors and $\Delta t$ is the time interval. 
The model gives $P_{i}(0) \simeq 4.1\times10^{-18}\:\mathrm{km}^{-2}\mathrm{y}^{-1}$ for both Ceres and Vesta. This is 
somewhat higher than the mean value $P_i = 2.85\times10^{-18}\:\mathrm{km}^{-2}\mathrm{y}^{-1}$ usually considered for 
the whole asteroid belt \citep{1992Icar...97..111F}. For Ceres, \citet{2016NatCo...712257M} found $P_i=3.55\times10^{-18}\:\mathrm{km}^{-2}\mathrm{y}^{-1}$,
which is more consistent with our $P_{i}(0)$. The small difference can be related to the fact that our model distribution 
of main belt asteroids is more concentrated towards smaller semimajor axes, because the model does not account the 
presence of large collisional families at $a\gtrsim3$ au (mainly the Themis, Hygiea and Eos families). The mean impact 
velocities computed from our model are in the range of 4.6 to 7 km~$\mathrm{s}^{-1}$ for the whole simulated time 
interval. They show a slightly decreasing trend with $t$ during the earliest stages as asteroid impactors on high-$e$ 
orbits are removed. The mean velocity at $T=0$ is in good agreement with the current value \citep{1994Icar..107..255B}.

\subsection{Size distribution\label{subsec:mpf}}
A general procedure to analytically estimate the MPF has been outlined in \citet{2009AJ....137.4936M}. A limitation 
of this procedure arises from uncertainties in modeling the processes of crater erasure such as, in particular, 
the obliteration of older and smaller craters by newer and larger ones. The crater erasure can be included in the 
MPF through a weight function, as explained in \citet{2006Icar..183...79O} and \citet{2009AJ....137.4936M}.
Here we instead develop a Monte Carlo approach to forward model the crater size distribution 
\citep[also see][]{2016NatCo...712257M}. 

To simulate the formation of craters we combine the observed size distribution of the main belt asteroids 
with the chronology functions obtained from our dynamical model. The size distribution is constructed from 
the WISE/NEOWISE observations \citep{2011ApJ...741...68M, 2016PDSS..247.....M}\footnote{Available
at the NASA PDS Small Bodies Node, \url{https://sbn.psi.edu/pds/resource/neowisediam.html}},
which is practically complete down to $D_{\mathrm{ast}}\simeq9$-10 km. For diameters slightly smaller than that, 
we adopt an extrapolation $\mathcal{N}=10^{\,\alpha}D_{\mathrm{ast}}^{\,\,\,\,-\gamma}$, where $\alpha=6.5,\,\gamma=-2.6$ for the 
distribution of the whole main belt, and $\alpha=6.23,\,\gamma=-2.54$ for the main belt background, i.e. subtracting 
the members of known asteroid families. These extrapolations were obtained by fitting the size distribution 
of asteroids slightly larger than 10 km by a power law and extending the power law below 10 km. 

Our model consists of the following steps:
\begin{enumerate}
\item We define the minimum impactor diameter, $D_{\mathrm{ast},0}$, that needs to be accounted for to match the smallest craters 
that we want to model.
\item We use Eq. (\ref{eq:ntd}) to determine the average number of impacts $\overline{n}_{\mathrm{imp}} =n(T,D_{\mathrm{ast},0})$ at $T=0$ Ga.
\item We draw the actual number of impacts $n_{\mathrm{imp}}$ over the desired time span from a Poisson distribution with mean $\overline{n}_{\mathrm{imp}}$.  
\item We generate $n_{\mathrm{imp}}$ craters from main belt impactors larger than $D_{\mathrm{ast},0}$ using the following procedure:
\begin{enumerate}
\item From the main belt size distribution, we draw the size $D_{\mathrm{ast}}$ of the impactor (in m).
\item From the chronology function, we draw the time $T$ that will represent the crater age.
\item We obtain the velocity $v$ of the impact (in $\mathrm{m\,s}{}^{-1}$) at the time $T$. 
Note that this is more accurate than just drawing a value from the overall impact velocity distribution, 
because velocities are  slightly higher at earlier times.
\item We set the impact angle $\theta=45^{\circ}$ \citep{1962pam..book.....K}.
\item We compute the crater diameter $D_{\mathrm{crat}}$ (in m) using the scaling law from \citet{2016Icar..271..350J}
for non-porous targets:
\begin{equation}
D_{\mathrm{crat}}=1.52\,D_{\mathrm{ast}}^{\,\,\,0.88}\,v^{0.5}\,\left(\sin\theta\right)^{0.38}\left(\frac{\delta}{\rho}\right)^{0.38}g^{-0.25}D_{\mathrm{sc}}^{\,\,-0.13}\ .\label{eq:scal}
\end{equation}
Here, $\delta$ is the impactor's density, $\rho$ is the target's density, $g$ is the target's surface gravity 
(in $\mathrm{m\,s}{}^{-2}$), and $D_{\mathrm{sc}}$ is the simple-to-complex transition diameter (i.e., the diameter 
for which the crater starts to develop complex structures, such as multiple ridges, concentric rings, etc.). The 
values of these parameters adopted here for Ceres and Vesta are given in Table \ref{params}.
\end{enumerate}
\item We assign to each crater the initial weight $W=1$. 
\item To account for crater erasure, we consider, one by one, the model-generated craters with size $D_{\mathrm{crat}}$ and age $T$.
We then select all craters with sizes $<D_{\mathrm{crat}}$ and ages $>T$, and subtract from their weights an amount $\pi D_{\mathrm{crat}}^{2}/(4S)$, which is the ratio of the crater surface area to the body surface area.
When $W$ becomes 0, the corresponding crater is assumed to be totally obliterated. This recipe is designed to model the crater overlap only, i.e. a ``cookie cutting'' approach.
\item The final size distribution of craters is obtained by adding all weights of craters with diameter $D_{\mathrm{crat}}$ 
together. 
\item The steps (3) to (7) are repeated 1000 times to build up statistics. We compute the mean and the $1\sigma$ 
uncertainty of crater size distributions and compare them to observations. 
\item Optionally, we can include formation of a large basin at a specific time. This is done, for example, to test 
the erasure of older and smaller craters by the Rheasilvia basin formation.
\end{enumerate}
The crater erasure algorithm in step (4) is a simple method that only accounts for crater overlap. It does not take
into account, for example, that the material ejected from large craters may degrade/bury small craters at considerable distance 
from the impact site. It is therefore expected that our method could underestimate the erasure of small craters. Here, however, we restrict our analysis to $D_{\mathrm{crat}}>50$ km craters for which this effect may not be important. 

\subsection{Caveats\label{sec:Caveats}}
As a by-product of the procedure outlined above we obtain a set of $D_{\mathrm{crat}}$ \textsl{vs.} $D_{\mathrm{ast}}$ values indicating that the
scaling law in Eq. (3) approximately follows a linear dependence $D_{\mathrm{crat}}\simeq f_{\mathrm{sl}}\times D_{\mathrm{ast}}$, where $f_{\mathrm{sl}}$ 
is a constant factor, at least in the 
size range considered here. The typical values of $f_{\mathrm{sl}}$ are in the range $\sim11$-13 for Ceres and 
$\sim8$-10 for Vesta. Therefore, if we want to fit the size distribution of craters with $D_{\mathrm{crat}} > 60$ km, we have 
to set $D_{\mathrm{ast},0}\sim6$ km in the case of Vesta and $D_{\mathrm{ast},0}\sim4$ km in the case of Ceres. This creates a problem because the dynamical model used here 
is strictly reliable only for $D_{\mathrm{ast}}\gtrsim 10$ km (because it does not account for the size-dependent processes such 
as the Yarkovsky effect or collisional fragmentation).  

The Yarkovsky drift of a $D_{\mathrm{ast}} = 4$ km asteroid is expected to be $\sim0.04$ au~Gy$^{-1}$. The drift may be directed 
inward or outward, depending on asteroid's spin axis orientation. The intrinsic collision probability of the target is not
expected to be significantly affected by this, because the inward and outward drifts would average out (assuming a random 
orientation of spin axes). The main effect of the Yarkovsky drift should consist in larger depletion of small 
main belt asteroids relative to our size-independent model where asteroid orbits are expected to be more stable. This could potentially 
mean that the chronology function would have a somewhat steeper dependence with time for $D_{\mathrm{ast}} < 10$~km than for 
$D_{\mathrm{ast}}>10$~km impactors. The investigation of this effect is left for future work.   

The effects of collisional grinding are difficult to estimate. The collisional grinding removes mass over time and thus 
reduces the population of small asteroids. This happens on the top of the dynamical depletion.  The general expectation 
is that the belt should evolve faster initially when it is still massive \citep{2005Icar..175..111B}. Recall that we anchor the 
results of our dynamical model to the {\it current} population of small asteroids. Thus, running the clock back in time, 
our model must underestimate the actual number of impacts (because it does not account for impactors that were 
collisionally eliminated). 

The formation of asteroids families over the age of the solar system contributes to enhance the two effects discussed above, but it has also another consequence. There are several large collisional families in the outer asteroid belt
(Themis, Hygiea and Eos families) and these families have many $D_{\mathrm{ast}} \sim 10$ km members (\citealp{2015aste.book..297N}; \citealp{2015PDSS..234.....N}). Including 
these bodies in our calibration effectively means that we assume that all these families existed for the whole duration 
of our simulation (i.e., formed 4.56 Ga), which is clearly not the case because, for example, the Eos family formed 
only $\sim1.3$ Ga \citep{2006Icar..182...92V}. To test how this approximation affects our results, we can 
remove asteroid families from the main belt and calibrate our chronology on the current main belt background. These 
tests show a variation in the number of impacts by a factor of $\sim2$. The uncertainty of our results, described below, 
cannot be better than that. 

Finally, another possible source of uncertainty is the contribution to the collisional rates in the main belt of the population of Hungaria asteroids, which may have constituted a significant early population depending on the eccentricity history of Mars \citep{2018Icar..304....9C}. Model CASE1B from \citet{2017AJ....153..103N} does account for a primordial population of asteroids in the range $1.6<a<2.1$ au, the so called E-belt \citep{2012Natur.485...78B}. Therefore, the derived production functions and chronologies used here include the effects of this population. However, model CASE1B did not reproduce well the currently observed population of Hungarias, because the E-belt became more depleted that it should, especially at later times \citep{2015AJ....150..186R}. In any case, the uncertainty introduced by this effect is small and would be within the factor of 2 discussed above.  

\section{Results\label{sec:Results}}
\subsection{Comparison of lunar and asteroid chronologies} 
The chronology functions obtained in our model for Vesta, Ceres and the Moon are compared in Fig. \ref{crono}. 
The lunar chronology shows a vast number of impacts during the early epochs when the impactor flux is at least 
$\sim2$ orders of magnitude higher than at the present time \citep{2017AJ....153..103N}. This happens because many
main belt asteroids become destabilized during the planetary migration/instability and evolve into the terrestrial 
planet region, which leads to a strong asteroid bombardment of the Moon and terrestrial planets. In contrast, 
the impact flux on Vesta and Ceres is more unchanging with time. This happens because Vesta and Ceres orbit within the main belt and are continuously impacted by asteroids. 
For them, the early bombardment is not as dramatic 
as for the Moon. This means that the lunar chronology does not apply to Vesta or Ceres. These considerations 
also imply that that Vesta's and Ceres's craters should be on average younger than the lunar craters.   

\citet{2014P&SS..103..131O} reached similar conclusions. To illustrate this, we show the Vesta chronology 
from \citet{2014P&SS..103..131O} in Fig. \ref{crono}b. We used Eqs. (16) and (18) in their paper and scaled 
their MPF (their figure 1) assuming a linear scaling law with $8\leq f_{\mathrm{sl}}\leq20$. Note that 
$f_{\mathrm{sl}}\sim9$ reproduce well the scaling law of \citet{2016Icar..271..350J} for Vesta. We would 
therefore expect that our results for Vesta should plot near the upper limit of their chronology function
range, and this is indeed the case. In \citet{2014P&SS..103..131O}, the Vesta's chronology was pieced together 
from several publications and was compared with the lunar chronology of \citet{2001SSRv...96...55N} (which
was obtained by yet another method). The advantage of our approach is that all chronologies are derived from 
a single, self-consistent physical model.   

\subsection{Impact flux for early and late instabilities}

\begin{figure*}
\caption{A comparison of model impact chronologies for Ceres (blue line), Vesta (red line) and the Moon (black line). 
In panel (a), the chronology functions were normalized to one asteroid remaining in the main belt today. 
The solid lines are the chronologies for the instability time $T_{\mathrm{inst}}=4.5$ Ga and the dashed lines are 
the chronologies for $T_{\mathrm{inst}}=3.9$ Ga. In (b), the functions are absolutely calibrated to 
$\mathcal{N}(>\!\!9\,\mathrm{km})\simeq10\,500$ main belt asteroids. The shaded area in (b) represents Vesta's 
chronologies obtained in \citet{2014P&SS..103..131O}, assuming $f_{\mathrm{sl}}=8$ (the upper boundary) and
$f_{\mathrm{sl}}=20$ (the lower boundary).}
\label{crono}
\centering{}\includegraphics[width=0.49\columnwidth]{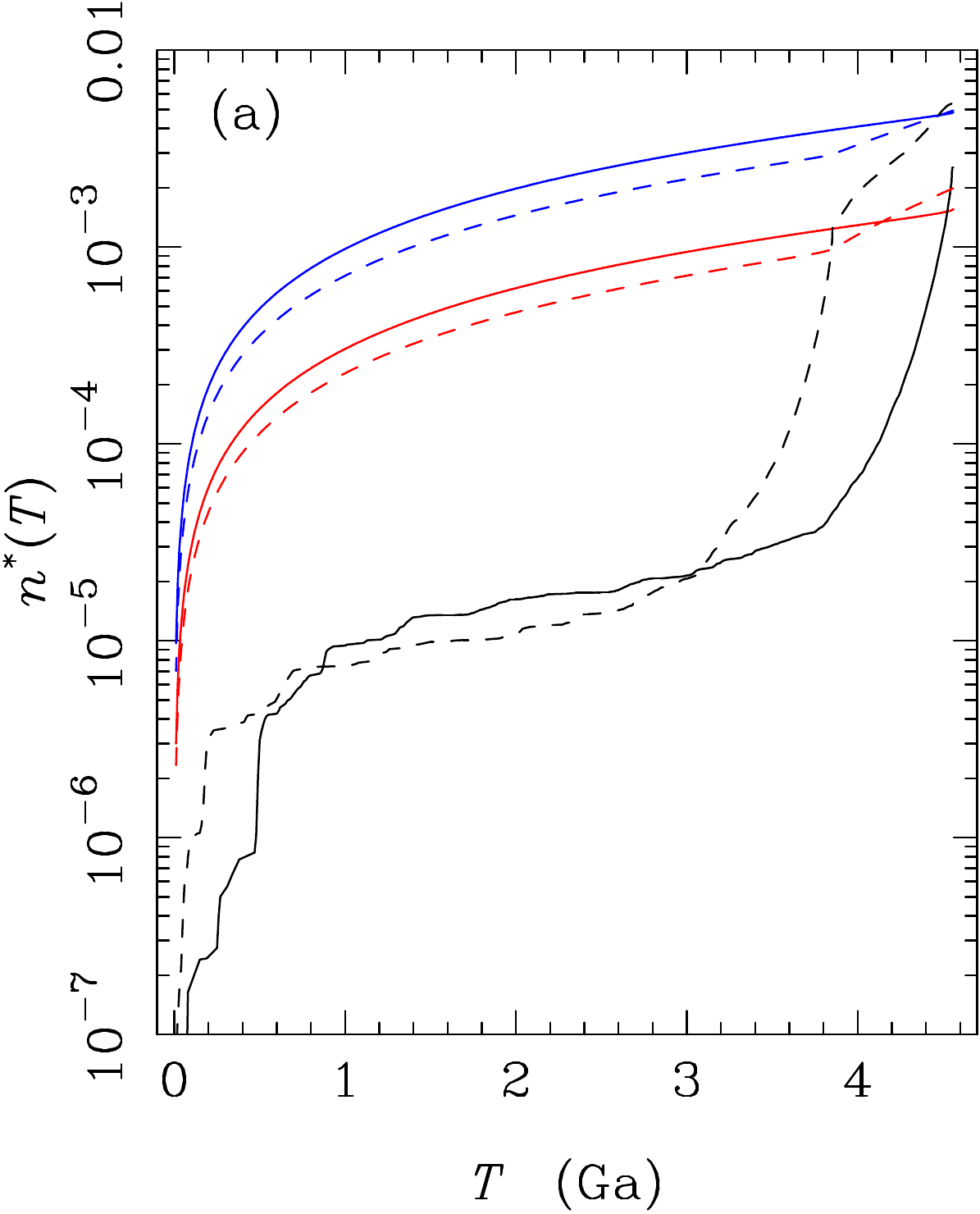}\includegraphics[width=0.49\columnwidth]{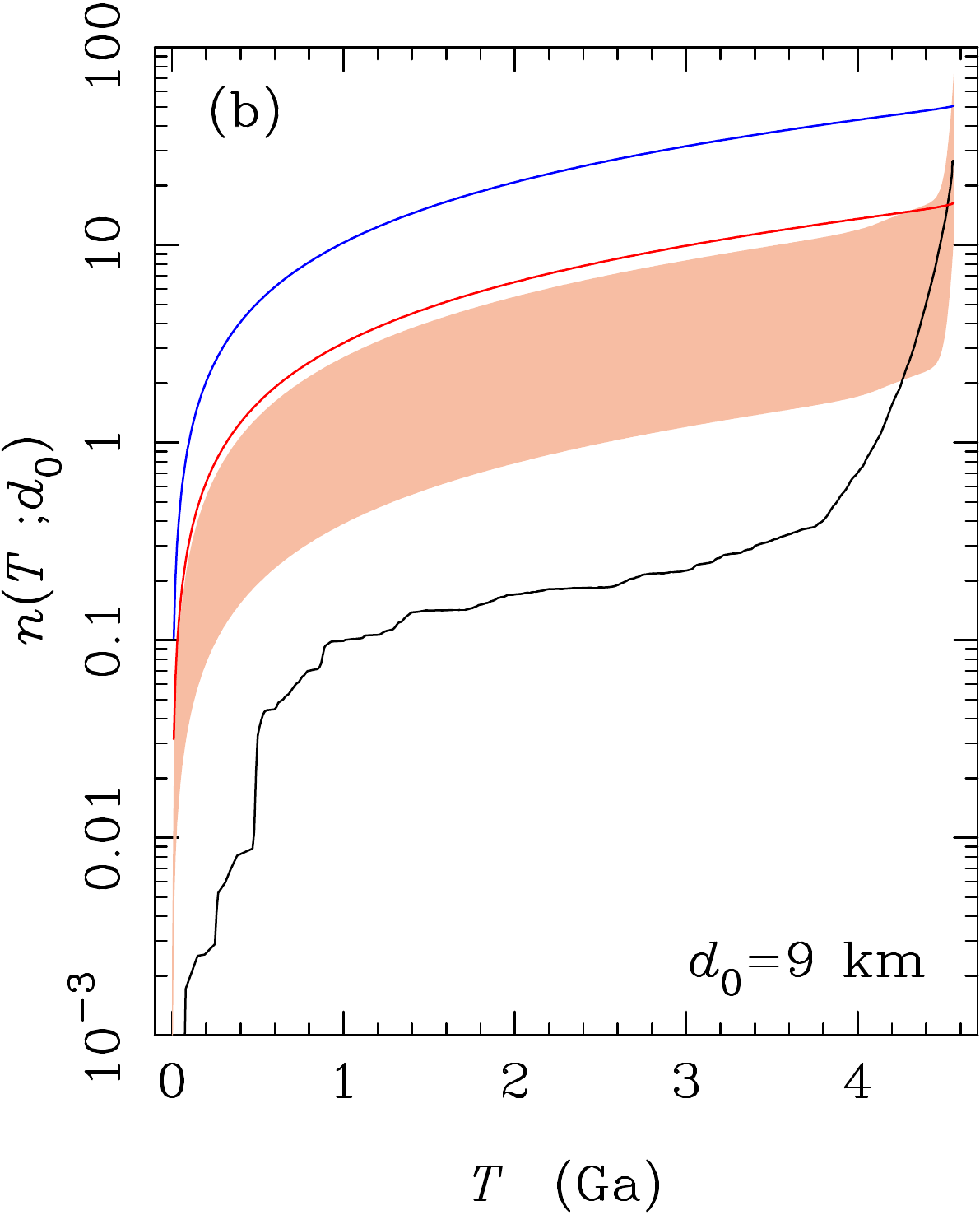}
\end{figure*}

The time of planetary migration/instability is a crucial parameter for the Moon as it substantially changes 
the lunar impact flux during early stages and the overall number of impacts (Fig. \ref{crono}a). Vesta's
and Ceres's impact records are much less sensitive to this parameter. Indeed, Fig. \ref{crono}a shows that
the records are nearly identical for $T_{\mathrm{inst}}=4.5$ Ga and $T_{\mathrm{inst}}=3.9$ Ga. We therefore do not
expect to find many clues about the LHB or the early evolution of the giant planets by analyzing the 
crater record of these asteroids. Given that other available constraints indicate that the instability
happened early \citep{2018NatAs...2..878N}, we give preference to the early instability case in the rest of the paper.
We find no appreciable difference for the Gaussian and Grand Tack initial distributions. The Gaussian 
initial distribution, as described in Sect. \ref{sec:Dynamical-model}, is used in the following analysis.  

The early instability model suggests that the Moon should have 
registered $\sim27$ impacts from $D_{\mathrm{ast}}>9$ km asteroids over the age of the solar system (see also \citealp{2017AJ....153..103N}),
while Ceres and Vesta registered $\sim51$ and $\sim16$ such impacts, respectively (Fig. \ref{crono}b). According to 
\citet{2014P&SS..103..131O}, Vesta would have registered between 10 and 75 impacts of $D_{\mathrm{ast}}>9$ km asteroids, but $\sim70$\% of these 
impacts would have occurred during the first 50 My of evolution. In general, O'Brien et al.'s chronology produces 
$\sim1.5$ times fewer impacts per Gy during the last $\sim4$ Gy than our chronology (assuming $f_{\mathrm{sl}}\sim9$). 
This discrepancy is, at least in part, related to the fact that O'Brien et al.'s chronology shows a drop at the 
very beginning, reflecting their attempt to account for strong depletion of the main asteroid belt by processes
not modeled here (e.g., planetary embryos, Grand 
Tack).\footnote{The strong 
depletion of the asteroid belt was thought to be needed, because the formation models based on the 
minimum mass solar nebula suggested that the primordial mass of the main belt was 100-1000 times larger than 
the present one \citep{1977Ap&SS..51..153W}. Also, the classical model of asteroid accretion by collisional 
coagulation required a large initial mass to produce 100-km class objects. The formation paradigm has shifted,
however, with more recent models favoring a low initial mass \citep{2015aste.book..493M}.}

\citet{2016NatCo...712257M} derived a chronology function for Ceres that has a very similar shape to 
O'Brien et al.'s chronology for Vesta. It also shows a drop during the first 50 My of evolution due
to a presumably strong primordial depletion of the main belt. Using this chronology, they predicted 180 and 
90 impacts from impactors with $D_{\mathrm{ast}}>10$ km and $D_{\mathrm{ast}}>13$ km, respectively. According to their scaling laws, 
these impactors produce craters with $D_{\mathrm{crat}}\sim100$ km. About 70\% of these impacts happen during the first
400 My of evolution (i.e. before the dynamical instability that they place at 4.1 Ga). Compared to that, 
our model implies $\sim$4 times fewer impacts and we do not find any significant difference between the 
number of impacts for the early and late instability cases. The number of craters of various sizes expected 
from our model is reported in Table \ref{tab-counts}. For Vesta, these numbers are in a good agreement 
with observations, especially if we account for modest crater erasure (see Sect. \ref{subsec:mpf}). For Ceres, strong crater erasure by viscous relaxation may be required (Sect. \ref{sec:Ceres-craters}).

\subsection{Vesta's craters}

\begin{figure*}
\caption{The simulated (solid lines) and observed (red dots) cumulative size distribution of Vesta's craters. 
The dashed lines are the $1\sigma$ uncertainties of model distributions obtained from 1000 trials with the 
Monte Carlo method described in Sect. \ref{subsec:mpf}. In each trial, we generate a number $n$ of impacts of main belt asteroids, where $n$ is drawn from a Poisson distribution with mean $\overline{n}=47$. This corresponds to the expected number of impacts of asteroids with $D_{\mathrm{ast}} \geq 6$ km (families included) over 4.56~Gy. The early instability chronology is used here. The error bars correspond to $\pm\sqrt{N}$, where $N$ is the number of craters identified on the surface 
\citep{2012Sci...336..690M}. From left to right, the panels show results without crater erasure, with crater
erasure, and with crater erasure and assuming that the Rheasilvia basin formed at 1 Ga.}
\label{distvesta}
\centering{}\includegraphics[width=0.99\textwidth]{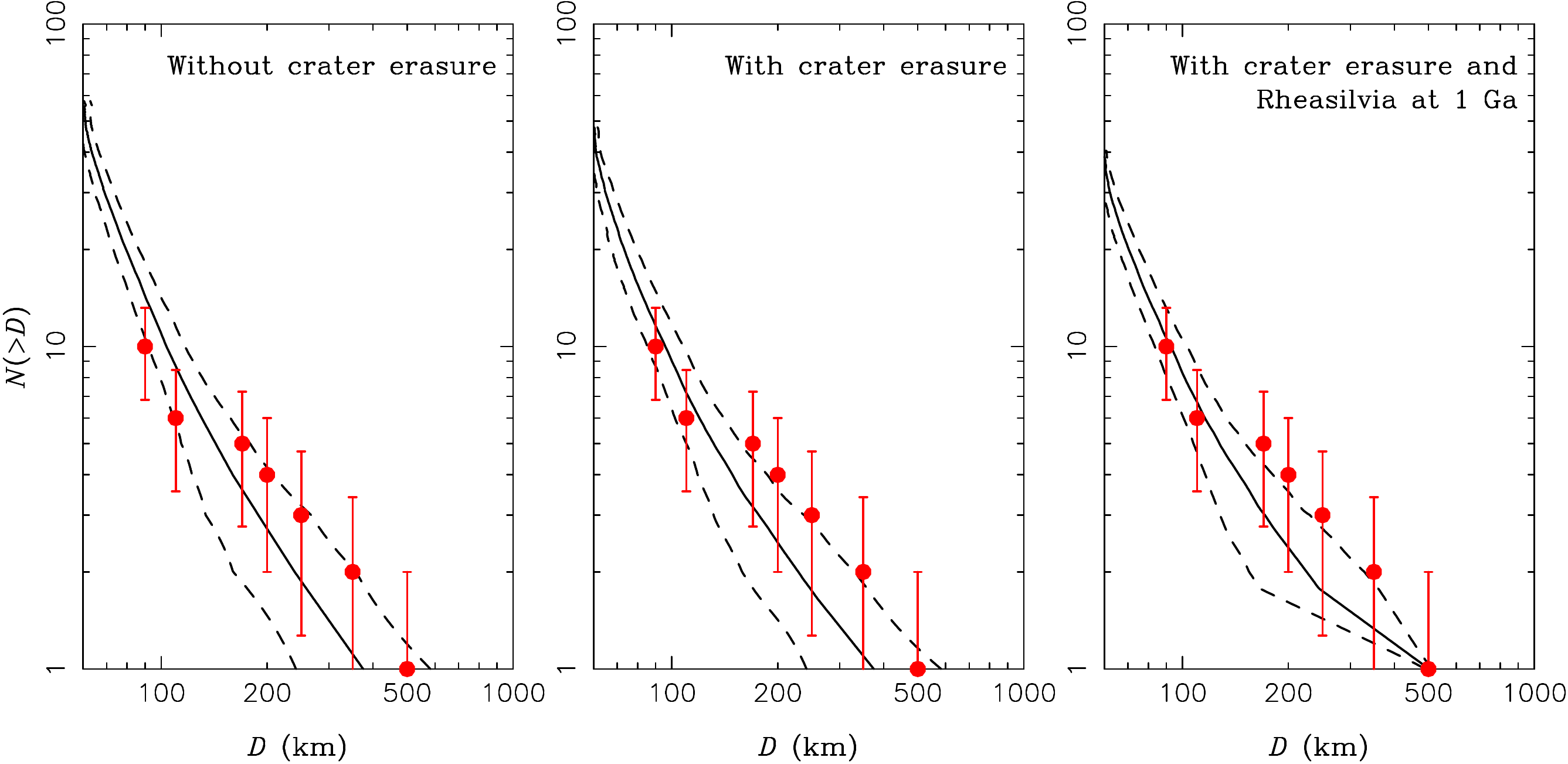}
\end{figure*}

Figure \ref{distvesta} compares our model size distributions of Vesta's craters to observations. 
To introduce this comparison, recall that we have blindly taken a dynamical model of the 
asteroid belt evolution (i.e., without any a priori knowledge of what implications the model will have for 
the Vesta's crater record) and used a standard scaling law to produce the crater record. There is not much 
freedom in this procedure. If the dynamical model were not accurate, for example, we could have obtained 
orders of magnitude more or less craters than what the {\it Dawn} mission found. But this is not the case. 
In fact, there is a very good general agreement between the model results and observations. This also 
shows that the caveats discussed in Sect. \ref{sec:Caveats} do not (strongly) influence the results. 

In more detail, in a model where no crater erasure is taken into account (left panel of Fig. 
\ref{distvesta}), the agreement is excellent for craters with $D_{\mathrm{crat}}>100$ km. There is a small difference 
for $D_{\mathrm{crat}}\lesssim100$ km, where the model distribution steeply raises and slightly overestimates the 
number of craters. A similar problem was identified in \citet{2014P&SS..103..131O}. We tested whether
this issue may be a result of crater erasure. Indeed, when crater erasure is included in the model 
(the middle panel of Fig. \ref{distvesta}), the size distribution shifts down and becomes slightly 
shallower. It now better fits the data in the whole range modeled here. The results do not change 
much when we include the presumed Rheasilvia basin formation at $\sim1$ Gy ago (right panel of Fig. 
\ref{distvesta}).\footnote{If the dynamical model is calibrated on the main belt background (i.e., 
asteroid families removed; Sect. \ref{sec:Caveats}), we obtain $\sim2$ times fewer craters. This does not make 
much of a difference on the logarithmic scale in Fig. \ref{distvesta}, but the overall fit without 
crater erasure becomes slightly better.}

In summary, our model works really well to reproduce the Vesta's crater record and a modest crater 
erasure may be needed to better fit the number of $D_{\mathrm{crat}} \lesssim 100$ km craters.

\subsection{Ceres's craters\label{sec:Ceres-craters}}
\begin{figure*}
\caption{The simulated (solid lines) and observed (blue dots) cumulative size distribution of Ceres's craters. 
The dashed lines are the $1\sigma$ uncertainties of model distributions obtained from 1000 trials with the 
Monte Carlo method described in Sect. \ref{subsec:mpf}. 
In each trial, we generate a number $n$ of impacts of main belt asteroids, where $n$ is drawn from a Poisson distribution with mean $\overline{n}=416$. This corresponds to the expected number of impacts of asteroids with $D_{\mathrm{ast}} \geq 4$ km (families included) over 4.56~Gy. The early instability chronology is used here.  
The error bars correspond to $\pm\sqrt{N}$, where $N$ is the number of craters identified on the surface 
\citep{2012Sci...336..690M}. From left to right, the panels show results without crater erasure, with crater
erasure, and with crater erasure and viscous relaxation. The Vendimia \textit{planitia} is shown here as a 
possible impact basin with $D_{\mathrm{crat}}\simeq800$ km.}
\label{distceres}
\centering{}\includegraphics[width=0.99\textwidth]{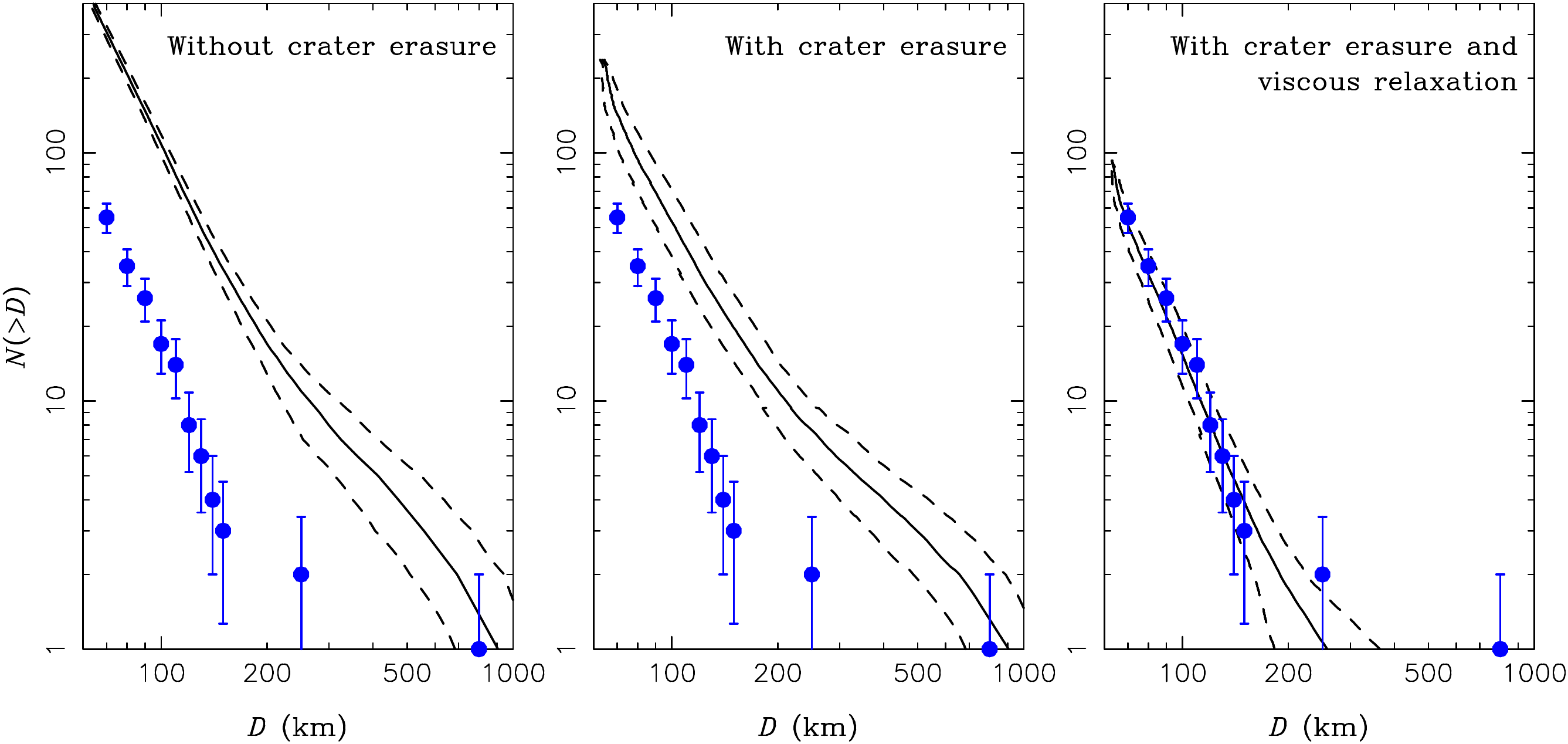}
\end{figure*}

Figure \ref{distceres} shows a similar comparison for Ceres. In this case, the model without crater 
erasure predicts nearly an order of magnitude more craters on Ceres's surface than the number of
actual craters. A similar problem was noted in \citet{2016NatCo...712257M}. The situation improves 
when the crater erasure is included in the model (middle panel of Fig. \ref{distceres}), but the problem
is not entirely resolved. We could have tried to erase craters more aggressively, for example, by assuming that small craters are degraded by distal ejecta from large craters \citep{2019Icar..326...63M}. But this would create problems for 
Vesta where the model with our conservative erasure method (craters must overlap to be erased) worked quite well.
Actually, \citet{2019Icar..326...63M} showed that crater degradation by energetic deposition of ejecta (e.g. secondary cratering/ballistic sedimentation) on the Moon works differently for the larger craters comparable to the crater sizes considered here \citep{2019EPSC...13.1065M}, so that mechanism would probably not be applicable in the cases of Ceres and Vesta.

Following \citet{2016NatCo...712257M}, we therefore investigate the effects of 
viscous relaxation (which are specific to ice-rich Ceres).
To empirically incorporate the effects of viscous relaxation in our model, we assume that the model 
weight of each crater diminishes according to the following prescription:
\begin{equation}
W=\exp\left(-T/\tau\right)\label{eq:ww}
\end{equation}
where e-folding timescale is a function of crater diameter,
\begin{equation}
\tau = C/ D_{\mathrm{crat}} \label{eq:tau}
\end{equation}
as supported by classical models of relaxation on icy surfaces \citep[e.g.][]{1973Icar...18..612J,2012GeoRL..3917204B,2013Icar..226..510B}. Here, $C = 4\pi\eta/\rho g$ is a constant depending on the viscosity $\eta$ of the surface layer.

The right panel of Fig. \ref{distceres} shows the model results for Ceres considering crater erasure together with viscous relaxation. In this case, we are able to fit the observed crater record assuming a value of $C\simeq 200$~km~Gy, which would imply a surface viscosity of $\sim 3\times 10^{23}$~Pa~s. This is about three orders of magnitude larger than the viscosity of pure ice at 180 K (the approximate temperature of Ceres surface), meaning that the particulate content volume in the icy surface layer needs to be significant. In fact, viscous relaxation of a purely icy surface is expected to be an aggressive process, with a typical e-folding timescale of only 1 My for the erasure of topographic wavelengths as short as 100 km. Our result is in line with more rigorous studies of the Ceres internal structure \citep{2017E&PSL.476..153F}, which infer a mechanically strong crust, with maximum effective viscosity $\sim 10^{25}$~Pa~s.

This gives some support to the viscous relaxation prescription discussed above. We caution, 
however, that the results are likely not unique and different combinations of crater erasure and viscous 
relaxation prescriptions (e.g., more aggressive crater erasure and longer viscous relaxation timescale)
could produce similarly good fits. 

In summary, we find that both erasure processes should be important for Ceres and $D_{\mathrm{crat}}\sim100$ km Ceres's craters should viscously relax on an e-folding timescale of $\sim 1$-2 Gy. This represents an interesting constrain on geophysical models of viscous relaxation and Ceres's composition. 

\subsection{Basins formation}
\begin{figure*}
\caption{The isochrones derived from our model production function for Vesta in panel (a), and Ceres in panel (b). Each line corresponds to a different age $T$. The production function for Vesta considers impactors with $D_{\mathrm{ast}}\geq 6$~km, while for Ceres it considers impactors with $D_{\mathrm{ast}}\geq 4$~km, in an early instability model. No crater erasure is used or accounted for in these plots.}
\label{isocro}
\centering{}\includegraphics[width=0.49\textwidth]{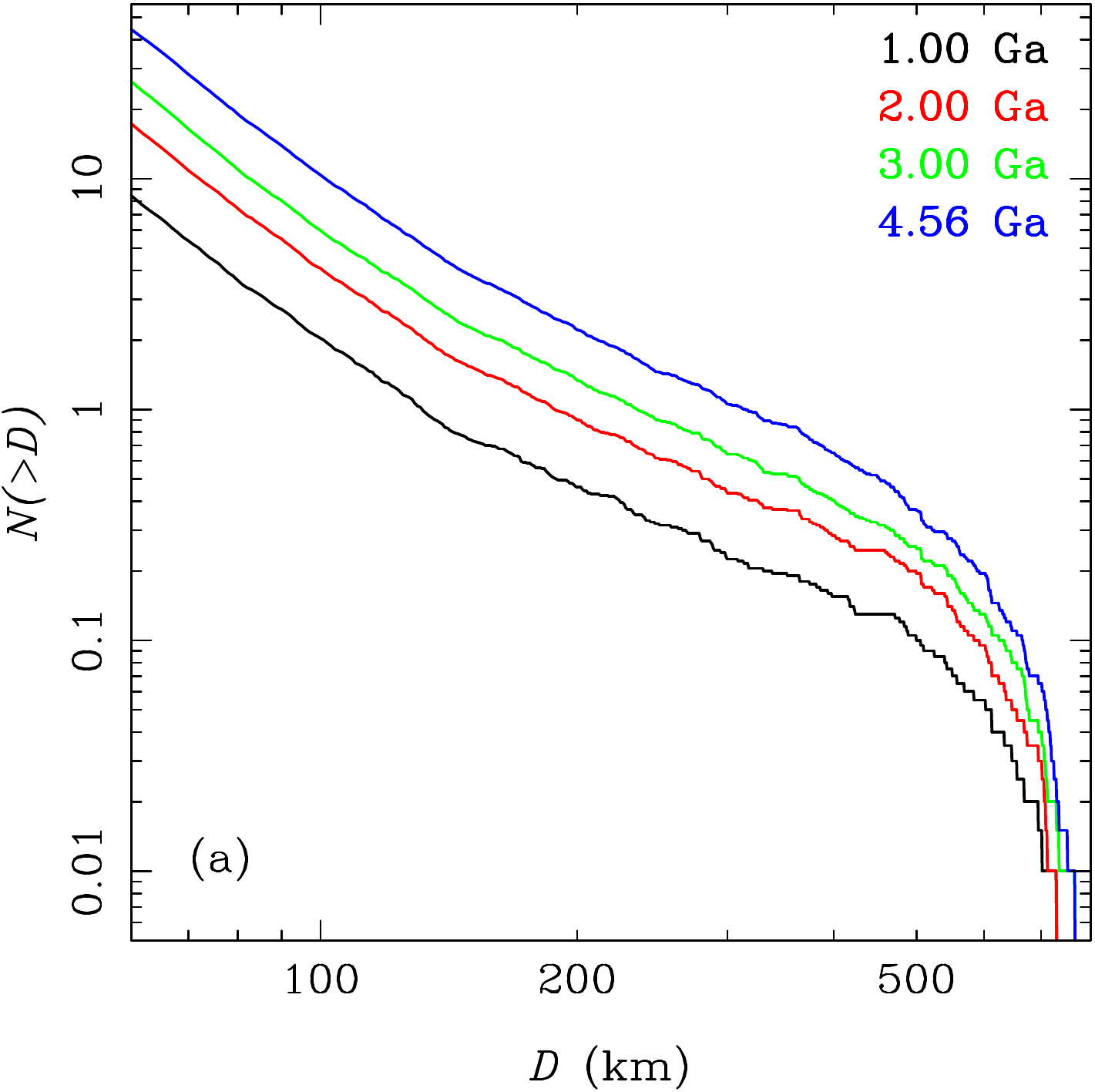}~~~\includegraphics[width=0.49\textwidth]{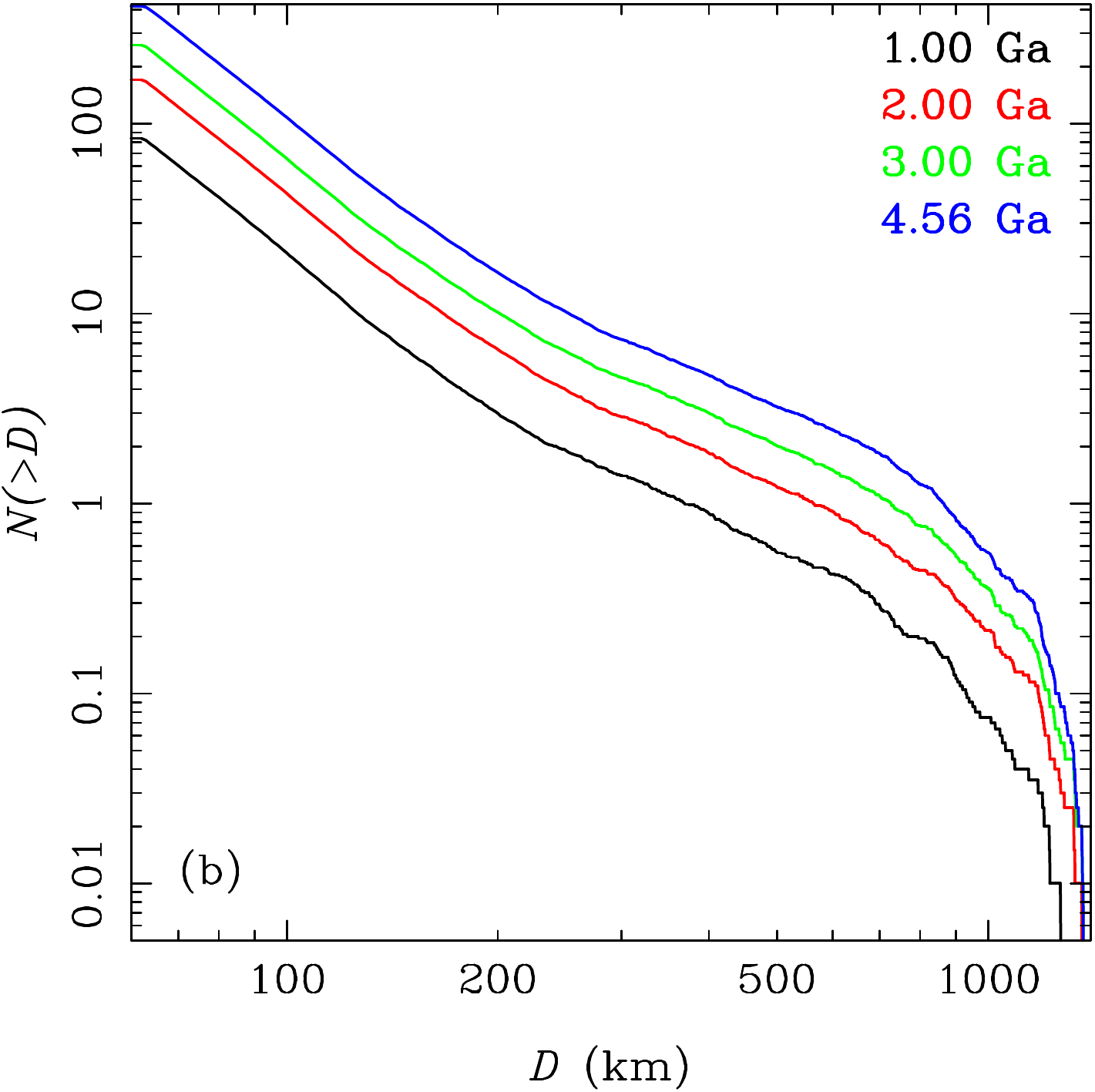}
\end{figure*}

Here, we discuss the probability of forming large craters or basins ($D_{\mathrm{crat}}>400$~km) on Vesta and Ceres at different 
times in the past. 
One possible approach to this problem consists in computing the so-called isochrones for each body, i.e. the crater production function at different times $T$. For a given diameter $D_{\mathrm{crat}}$, each isochrone gives the expected number of craters $\mu(>D_{\mathrm{crat}},\,<T)$, and the probability of forming exactly $N$ (and only $N$) craters $>D_{\mathrm{crat}}$ in a time $<T$ is obtained from a Poisson distribution:
\begin{equation}
p_{\mu}(N)=\frac{\mu^N\,e^{-\mu}}{N!}\label{eq:pn}
\end{equation}
Figure \ref{isocro} shows the isochrones for Ceres and Vesta, as determined from our model, without considering any crater erasure. If we take the case of a 500 km basin on Vesta, we find that the expected value for the $T=1$~Ga isochrone is $\mu =0.10$, and from Eq. (\ref{eq:pn}) the probability of forming one basin is 9\%, while the probability of forming two basins is much smaller, 0.5\%. However, if we consider the $T=4.56$~Ga isochrone, the probability of forming two basins increases to 4.6\%. We recall that the probability of forming at least one 500 km basin in the last 1 Gy can be obtained as $1-p_{\mu}(0)$, which in this case would give a value of 9.5\%. Table \ref{tab-poison} summarizes the results for $D_{\mathrm{crat}}>400$ km.

\begin{figure*}
\caption{The probability of forming Vesta craters/basins at different times during the solar system history. 
In panel (a), we plot the probability of creating at least one crater larger than a given size since 
time $T$ ago. In (b), we plot the probability of creating at least two craters. Each line corresponds to 
a different cutoff size. The early instability is assumed here, and no crater erasure is included.}
\label{probab}
\centering{}\includegraphics[width=0.49\textwidth]{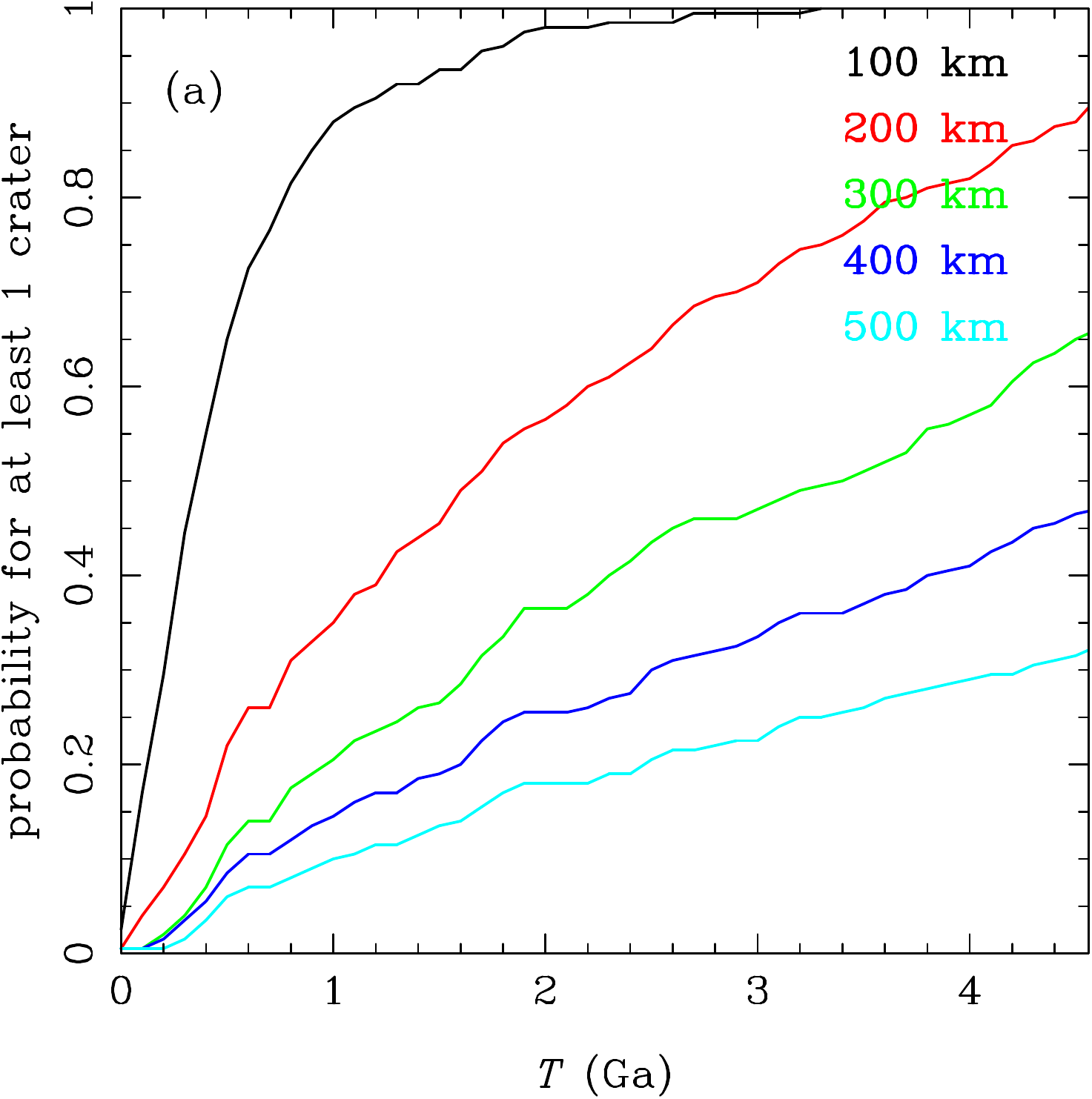}~~~\includegraphics[width=0.49\textwidth]{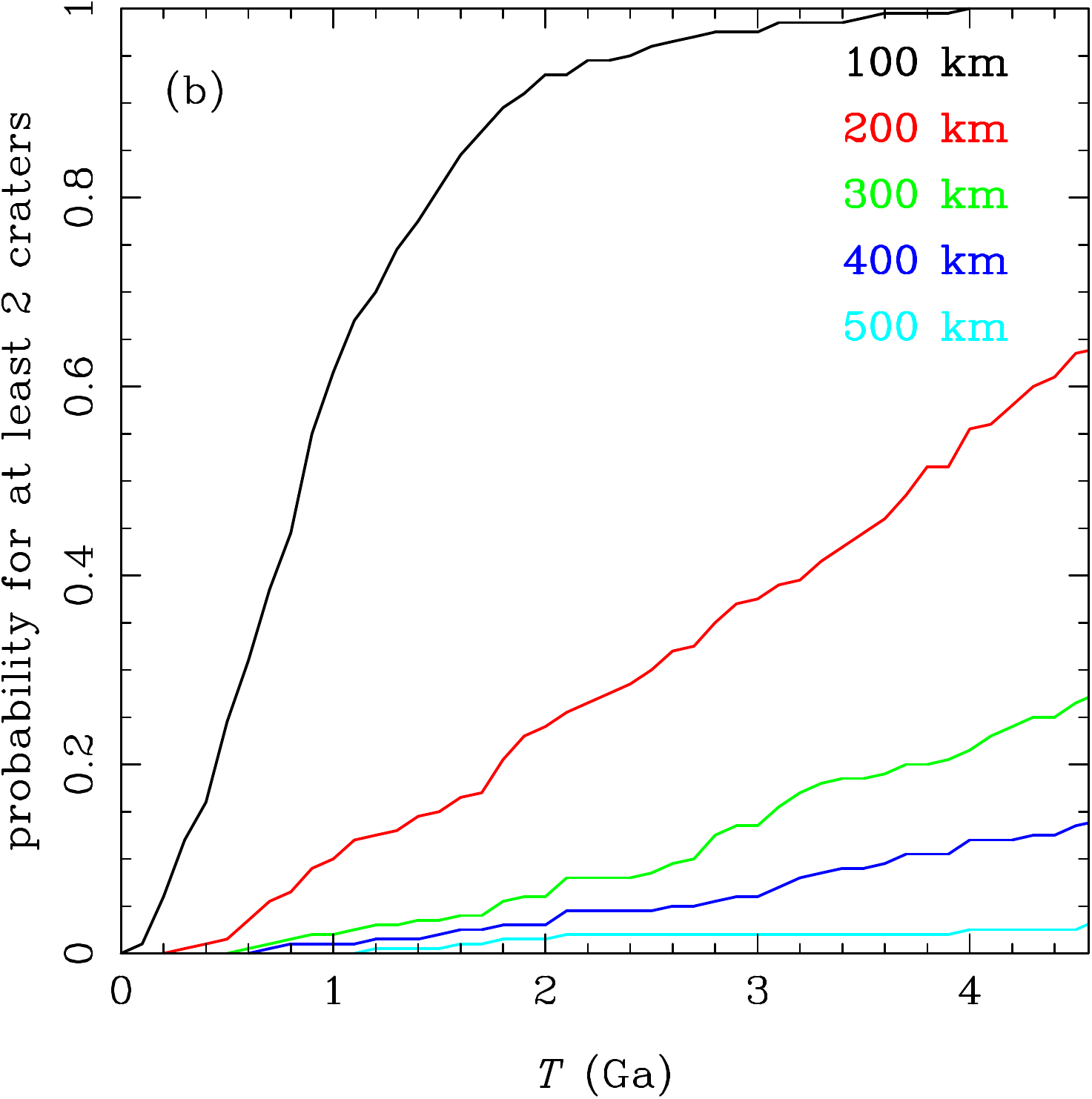}
\end{figure*}

Another possible approach consists in using our model to directly determine the probability of producing at least $N$ craters larger than a given size over a certain time span. This approach differs from
the previous one in that it does not rely on the Poisson statistics, but on the output of the Monte Carlo simulations. Figure \ref{probab} shows the probability of creating at least one crater (panel a) 
and at least two craters (panel b) larger than a cutoff diameter on Vesta. Again, no crater erasure is 
considered here. We find that the probability of creating the Rheasilvia basin with $D_{\mathrm{crat}}\simeq500$~km
(the cyan line in panel a) in the last 1 Gy (or 2 Gy) is 10\% (or 18\%). This is about 2.5 times more likely than
the probability reported in \citet{2014P&SS..103..131O}. This happens because our chronology function 
leaves more space for a relatively late formation of craters/basins. \citet{2014P&SS..103..131O}, 
instead, adopted a strong primordial depletion and had more basins forming early on (e.g. 
\citealp{2007Icar..191..434O}). If we consider $D_{\mathrm{crat}}>400$~km (blue line in panel a), the probabilities of forming at least one crater become 14\% in the last 1 Gy, and 25\% in the last 2 Gy. These values are slightly larger than those reported in Table \ref{tab-poison}, because the Poisson statistics constrains the formation of exactly $N$ craters.

The probability of forming both the Rheasilvia and Veneneia basins (the blue line in Fig. \ref{probab}b corresponding to $D_{\mathrm{crat}}=400$ km) is 15\% over the age of the solar system, and 6\% in the last 3 Ga. Again, these value are slightly larger than those reported in Table \ref{tab-poison}.
Table \ref{tab-bene} reports different probabilities assuming that the age of Rheasilvia is $\leq1$ 
Gy (note that we do not claim that this is an accurate age of the Rheasilvia basin; we merely test this 
assumption) and Veneneia is $>1$ Gy, for the models with early and late instabilities. The probabilities 
are slightly higher in the early instability model simply because, in this model, the rate of impacts is slightly 
higher in the past 1 Gy. Thus, a young age for Rheasilvia could potentially be more consistent with
an early instability model. In any case, our chronology still implies that most Vesta's craters/basins
should have preferentially formed early in the solar system history. 

\begin{figure}
\caption{The probability of forming Ceres's craters/basins at different times during the solar system history. 
The plot shows the probability of creating at least one crater larger than a given size since 
time $T$ ago. Each line corresponds to a different cutoff size. The early instability is assumed here, and no crater erasure is included.}
\label{probceres}
\centering{}\includegraphics[width=0.5\columnwidth]{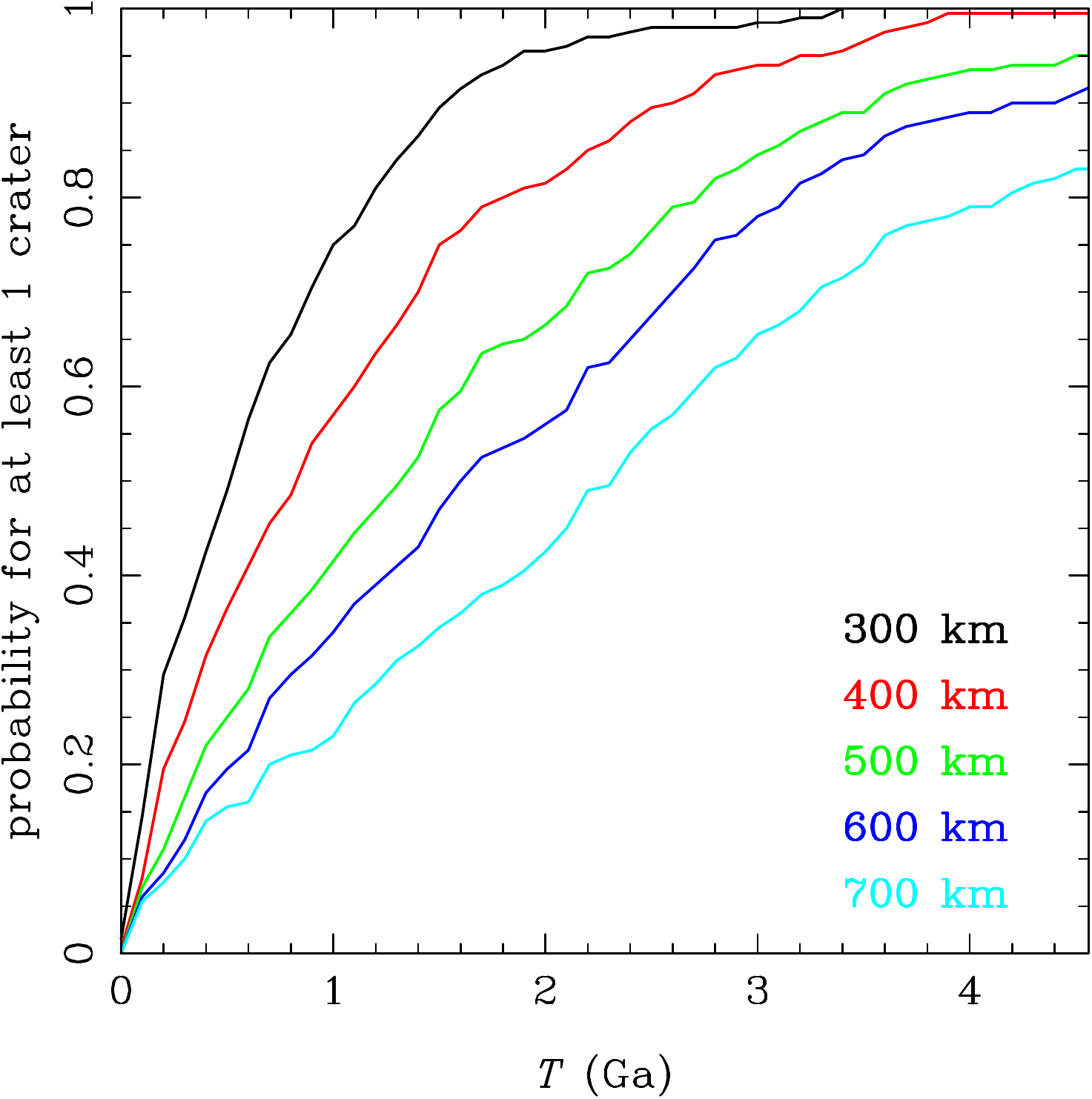}
\end{figure}

\begin{figure*}
\caption{The results of two Monte Carlo simulations of craters with $D_{\mathrm{crat}}>60$ km on Ceres (left) and Vesta (right). The simulation at the top illustrates a very unlikely case (less than 0.1\% of probability) where no large basin formed on Ceres, whereas two basins with $D_{\mathrm{crat}}>400$ km formed on Vesta. The simulation at the bottom shows a typical outcome of our model, with some very large young basins forming on Ceres. The craters/basins are projected on the surface using the McBryde-Thomas Flat Polar Quartic projection.}
\label{proyect}
\centering{}\includegraphics[width=0.49\textwidth]{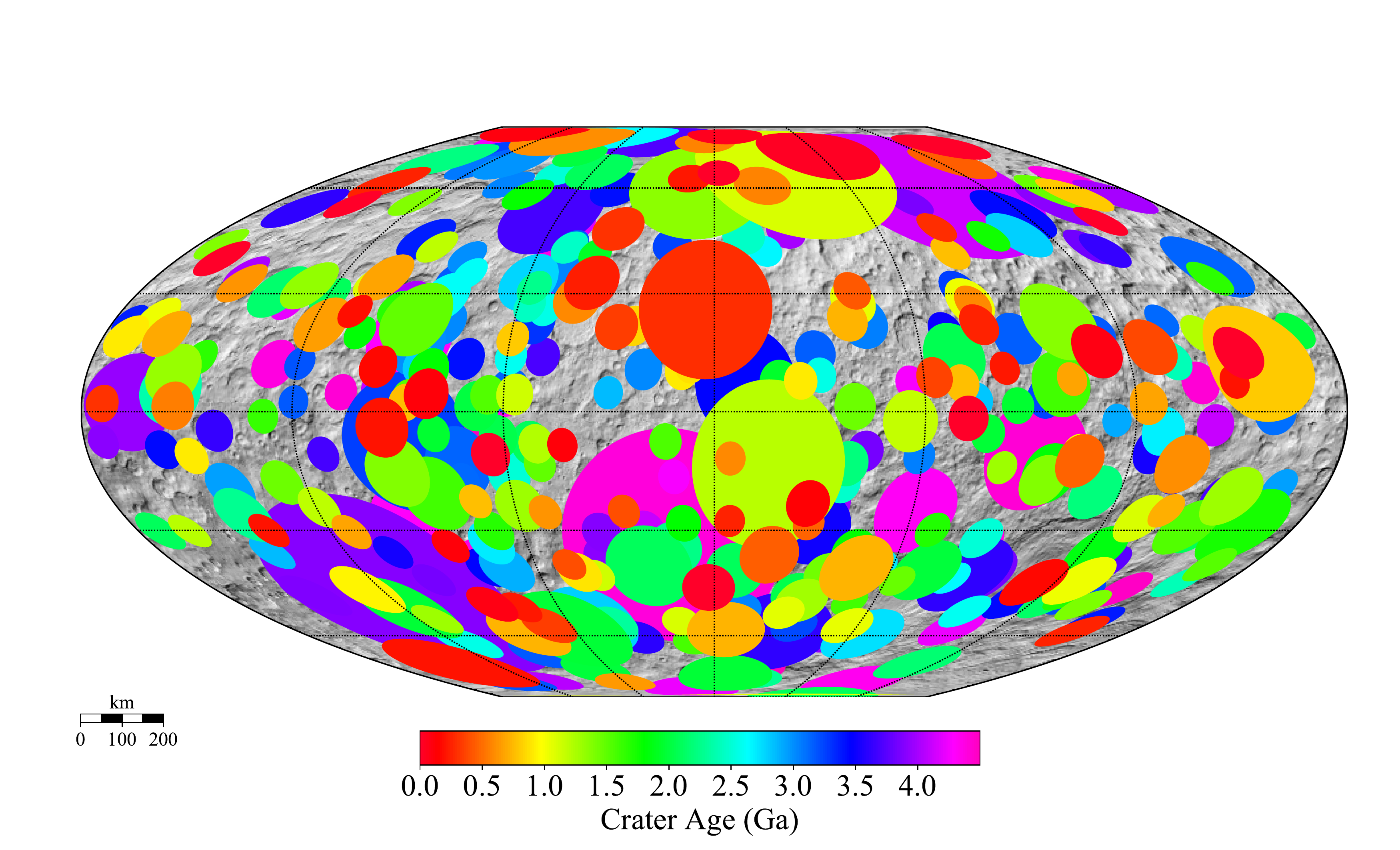},\includegraphics[width=0.49\textwidth]{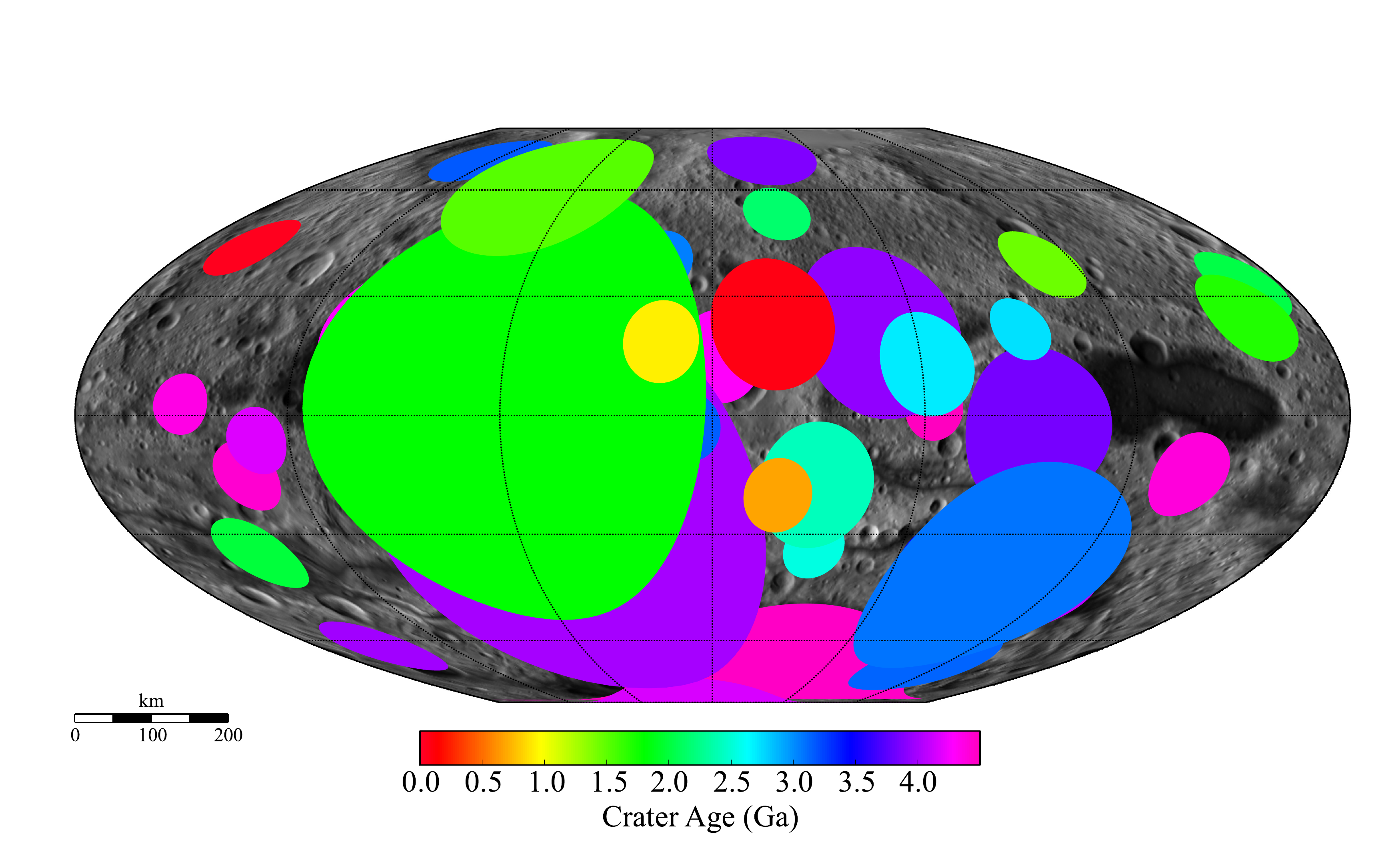}
\centering{}\includegraphics[width=0.49\textwidth]{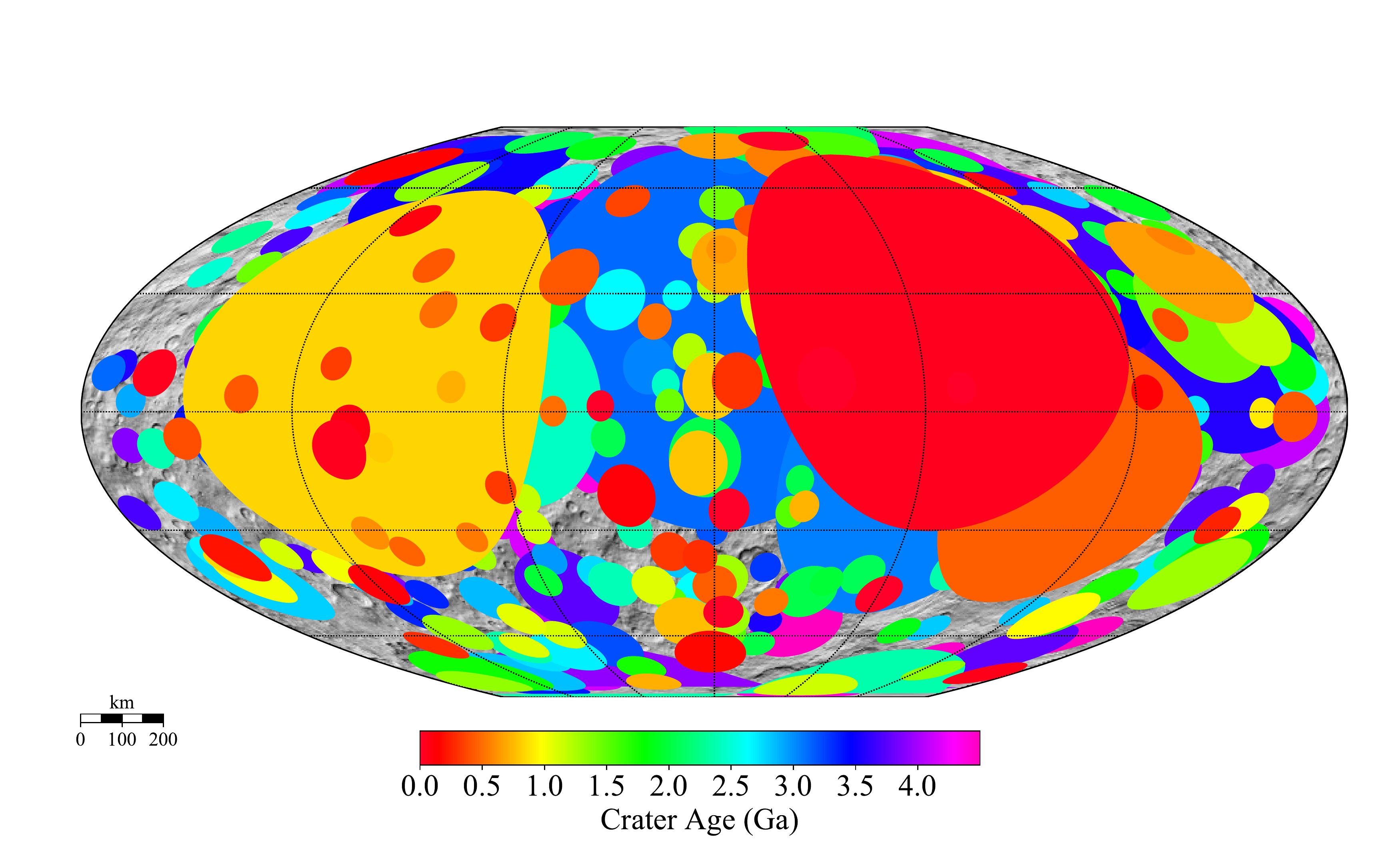},\includegraphics[width=0.49\textwidth]{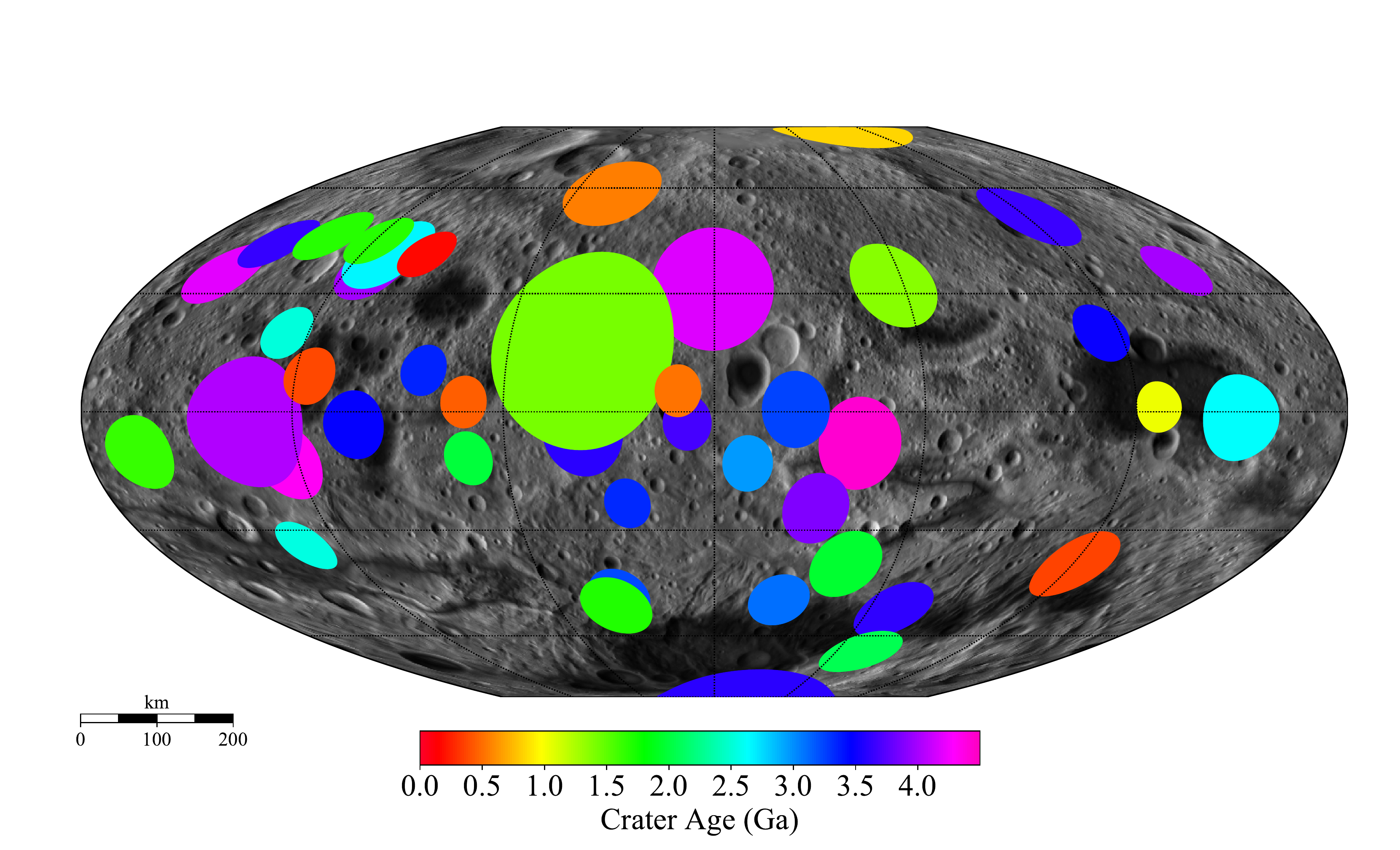}
\end{figure*}

Figure \ref{probceres} shows the results for Ceres. In this case, the probability of {\it not} creating any
basin with $D_{\mathrm{crat}}>400$ km over the age of the solar system is only 1\% (the red line in Fig. \ref{probceres}).  
Combining this result with the one for Vesta (see above) we estimate that the joint probability of creating
two $D_{\mathrm{crat}} > 400$~km basins on Vesta younger than 3 Gy and no $D_{\mathrm{crat}} > 400$ km basin on Ceres is only $<0.1$\%. 
Figure \ref{proyect} shows, at the top, the one exceptional case we found over 1000 realizations that fulfills the above condition. For comparison, an example of the typical outcome of our Monte Carlo model is shown at the bottom. This result emphasizes the need for efficient removal of Ceres's basins by viscous relaxation (or some other process).

\section{Conclusions\label{sec:Conclusions}}
Our findings can be summarized as follows:
\begin{itemize}
\item The crater chronologies of Ceres and Vesta are very different from that of the Moon. This is a consequence 
of the fact that both Vesta and Ceres spent their whole lifetimes in the asteroid belt and are impacted
all the time, whereas the Moon experienced a more intense bombardment during the first $\sim1$ Gy. 
This means that using the lunar chronology for Ceres and Vesta is incorrect. The scaled lunar chronology would 
imply that Vesta's basins must have formed very early in the solar system history, which may not necessarily 
be the case. 
\item Our crater chronologies of Ceres and Vesta are similar to those obtained in some previous studies 
(\citealp{2014P&SS..103..131O}; \citealp{2016NatCo...712257M}). In our chronology, however, the crater ages are 
not as concentrated toward the early times as in these works, allowing more impacts in the past 3~Gy.
\item The model crater record of Vesta matches observations (e.g., 10 known craters with $D_{\mathrm{crat}}>90$ km).
The model with crater erasure overpredicts, by a factor of $\sim3$, the number of $D_{\mathrm{crat}}>90$ km craters observed on the Ceres's surface. An additional erasure process such as, for example, the size-dependent viscous relaxation of craters (with $\sim 2$ Gy timescale for $D_{\mathrm{crat}}=100$ km craters), may be responsible for this discrepancy. 
\item We estimate that the probability of creating the Rheasilvia and Veneneia basins ($D_{\mathrm{crat}} >400$ km) on Vesta during the last 
3 Gy is $\simeq 6$\%, somewhat larger than found in the previous studies. A recent formation of the Rheasilvia basin can be more easily accepted in a dynamical model with the early instability, where the impact probabilities in the last 1 Gy are higher.
\item The probability of producing two large basins ($D_{\mathrm{crat}} > 400$ km) on Vesta and simultaneously not producing 
any basin on Ceres is interestingly small ($<0.1$\%). The relative paucity of large craters/basins on Ceres may 
be explained in a model with crater erasure and viscous relaxation.
\end{itemize}

\acknowledgments

The authors wish to thank David Minton for helpful comments and suggestions during the revision of this paper. FR's work was supported by the Brazilian National Council of Research (CNPq). DN's work was supported by the NASA SSERVI and SSW programs. 
The simulations were performed on the SDumont computer cluster of the Brazilian System of High Performance Processing (SINAPAD).

\bibliography{vesta}{}

\begin{thebibliography}{}
\expandafter\ifx\csname natexlab\endcsname\relax\def\natexlab#1{#1}\fi
\providecommand{\url}[1]{\href{#1}{#1}}
\providecommand{\dodoi}[1]{doi:~\href{http://doi.org/#1}{\nolinkurl{#1}}}
\providecommand{\doeprint}[1]{\href{http://ascl.net/#1}{\nolinkurl{http://ascl.net/#1}}}
\providecommand{\doarXiv}[1]{\href{https://arxiv.org/abs/#1}{\nolinkurl{https://arxiv.org/abs/#1}}}

\bibitem[{{Binzel} \& {Xu}(1993)}]{1993Sci...260..186B}
{Binzel}, R.~P., \& {Xu}, S. 1993, Science, 260, 186,
  \dodoi{10.1126/science.260.5105.186}

\bibitem[{{Bland}(2013)}]{2013Icar..226..510B}
{Bland}, M.~T. 2013, \icarus, 226, 510, \dodoi{10.1016/j.icarus.2013.05.037}

\bibitem[{{Bland} {et~al.}(2012){Bland}, {Singer}, {McKinnon}, \&
  {Schenk}}]{2012GeoRL..3917204B}
{Bland}, M.~T., {Singer}, K.~N., {McKinnon}, W.~B., \& {Schenk}, P.~M. 2012,
  \grl, 39, L17204, \dodoi{10.1029/2012GL052736}

\bibitem[{{Bottke} {et~al.}(2005){Bottke}, {Durda}, {Nesvorn{\'y}}, {Jedicke},
  {Morbidelli}, {Vokrouhlick{\'y}}, \& {Levison}}]{2005Icar..175..111B}
{Bottke}, W.~F., {Durda}, D.~D., {Nesvorn{\'y}}, D., {et~al.} 2005, \icarus,
  175, 111, \dodoi{10.1016/j.icarus.2004.10.026}

\bibitem[{{Bottke} {et~al.}(1994){Bottke}, {Nolan}, {Greenberg}, \&
  {Kolvoord}}]{1994Icar..107..255B}
{Bottke}, W.~F., {Nolan}, M.~C., {Greenberg}, R., \& {Kolvoord}, R.~A. 1994,
  \icarus, 107, 255, \dodoi{10.1006/icar.1994.1021}

\bibitem[{{Bottke} {et~al.}(2012){Bottke}, {Vokrouhlick{\'y}}, {Minton},
  {Nesvorn{\'y}}, {Morbidelli}, {Brasser}, {Simonson}, \&
  {Levison}}]{2012Natur.485...78B}
{Bottke}, W.~F., {Vokrouhlick{\'y}}, D., {Minton}, D., {et~al.} 2012, \nat,
  485, 78, \dodoi{10.1038/nature10967}

\bibitem[{{Carruba} {et~al.}(2005){Carruba}, {Michtchenko}, {Roig},
  {Ferraz-Mello}, \& {Nesvorn{\'y}}}]{2005A&A...441..819C}
{Carruba}, V., {Michtchenko}, T.~A., {Roig}, F., {Ferraz-Mello}, S., \&
  {Nesvorn{\'y}}, D. 2005, \aap, 441, 819, \dodoi{10.1051/0004-6361:20053355}

\bibitem[{{{\'C}uk} \& {Nesvorn{\'y}}(2018)}]{2018Icar..304....9C}
{{\'C}uk}, M., \& {Nesvorn{\'y}}, D. 2018, \icarus, 304, 9,
  \dodoi{10.1016/j.icarus.2017.04.015}

\bibitem[{{Farinella} \& {Davis}(1992)}]{1992Icar...97..111F}
{Farinella}, P., \& {Davis}, D.~R. 1992, \icarus, 97, 111,
  \dodoi{10.1016/0019-1035(92)90060-K}

\bibitem[{{Fu} {et~al.}(2017){Fu}, {Ermakov}, {Marchi}, {Castillo-Rogez},
  {Raymond}, {Hager}, {Zuber}, {King}, {Bland }, {Cristina De Sanctis},
  {Preusker}, {Park}, \& {Russell}}]{2017E&PSL.476..153F}
{Fu}, R.~R., {Ermakov}, A.~I., {Marchi}, S., {et~al.} 2017, Earth and Planetary
  Science Letters, 476, 153, \dodoi{10.1016/j.epsl.2017.07.053}

\bibitem[{{Hiesinger} {et~al.}(2016){Hiesinger}, {Marchi}, {Schmedemann},
  {Schenk}, {Pasckert}, {Neesemann}, {O'Brien}, {Kneissl}, {Ermakov}, {Fu},
  {Bland}, {Nathues}, {Platz}, {Williams}, {Jaumann}, {Castillo-Rogez},
  {Ruesch}, {Schmidt}, {Park}, {Preusker}, {Buczkowski}, {Russell}, \&
  {Raymond}}]{2016Sci...353.4759H}
{Hiesinger}, H., {Marchi}, S., {Schmedemann}, N., {et~al.} 2016, Science, 353,
  aaf4758, \dodoi{10.1126/science.aaf4759}

\bibitem[{{Holsapple} \& {Housen}(2007)}]{2007Icar..187..345H}
{Holsapple}, K.~A., \& {Housen}, K.~R. 2007, \icarus, 187, 345,
  \dodoi{10.1016/j.icarus.2006.08.029}

\bibitem[{{Johnson} {et~al.}(2016){Johnson}, {Collins}, {Minton}, {Bowling},
  {Simonson}, \& {Zuber}}]{2016Icar..271..350J}
{Johnson}, B.~C., {Collins}, G.~S., {Minton}, D.~A., {et~al.} 2016, \icarus,
  271, 350, \dodoi{10.1016/j.icarus.2016.02.023}

\bibitem[{{Johnson} \& {McGetchin}(1973)}]{1973Icar...18..612J}
{Johnson}, T.~V., \& {McGetchin}, T.~R. 1973, \icarus, 18, 612,
  \dodoi{10.1016/0019-1035(73)90064-X}

\bibitem[{{Mainzer} {et~al.}(2019){Mainzer}, {Bauer}, {Cutri}, {Grav},
  {Kramer}, {Masiero}, {Sonnett}, \& {Wright}}]{2016PDSS..247.....M}
{Mainzer}, A.~K., {Bauer}, J.~M., {Cutri}, R.~M., {et~al.} 2019, NASA Planetary
  Data System, urn:nasa:pds:neowise\_diameters\_albedos::2.0

\bibitem[{{Marchi} {et~al.}(2009){Marchi}, {Mottola}, {Cremonese}, {Massironi},
  \& {Martellato}}]{2009AJ....137.4936M}
{Marchi}, S., {Mottola}, S., {Cremonese}, G., {Massironi}, M., \& {Martellato},
  E. 2009, \aj, 137, 4936, \dodoi{10.1088/0004-6256/137/6/4936}

\bibitem[{{Marchi} {et~al.}(2010){Marchi}, {Barbieri}, {K{\"u}ppers},
  {Marzari}, {Davidsson}, {Keller}, {Besse}, {Lamy}, {Mottola}, {Massironi}, \&
  {Cremonese}}]{2010P&SS...58.1116M}
{Marchi}, S., {Barbieri}, C., {K{\"u}ppers}, M., {et~al.} 2010, \planss, 58,
  1116, \dodoi{10.1016/j.pss.2010.03.017}

\bibitem[{{Marchi} {et~al.}(2012{\natexlab{a}}){Marchi}, {Massironi},
  {Vincent}, {Morbidelli}, {Mottola}, {Marzari}, {K{\"u}ppers}, {Besse},
  {Thomas}, {Barbieri}, {Naletto}, \& {Sierks}}]{2012P&SS...66...87M}
{Marchi}, S., {Massironi}, M., {Vincent}, J.-B., {et~al.} 2012{\natexlab{a}},
  \planss, 66, 87, \dodoi{10.1016/j.pss.2011.10.010}

\bibitem[{{Marchi} {et~al.}(2012{\natexlab{b}}){Marchi}, {McSween}, {O'Brien},
  {Schenk}, {De Sanctis}, {Gaskell}, {Jaumann}, {Mottola}, {Preusker},
  {Raymond}, {Roatsch}, \& {Russell}}]{2012Sci...336..690M}
{Marchi}, S., {McSween}, H.~Y., {O'Brien}, D.~P., {et~al.} 2012{\natexlab{b}},
  Science, 336, 690, \dodoi{10.1126/science.1218757}

\bibitem[{{Marchi} {et~al.}(2016){Marchi}, {Ermakov}, {Raymond}, {Fu},
  {O'Brien}, {Bland}, {Ammannito}, {de Sanctis}, {Bowling}, {Schenk}, {Scully},
  {Buczkowski}, {Williams}, {Hiesinger}, \& {Russell}}]{2016NatCo...712257M}
{Marchi}, S., {Ermakov}, A.~I., {Raymond}, C.~A., {et~al.} 2016, Nature
  Communications, 7, 12257, \dodoi{10.1038/ncomms12257}

\bibitem[{{Marzari} {et~al.}(1996){Marzari}, {Cellino}, {Davis}, {Farinella},
  {Zappala}, \& {Vanzani}}]{1996A&A...316..248M}
{Marzari}, F., {Cellino}, A., {Davis}, D.~R., {et~al.} 1996, \aap, 316, 248

\bibitem[{{Masiero} {et~al.}(2011){Masiero}, {Mainzer}, {Grav}, {Bauer},
  {Cutri}, {Dailey}, {Eisenhardt}, {McMillan}, {Spahr}, {Skrutskie}, {Tholen},
  {Walker}, {Wright}, {DeBaun}, {Elsbury}, {Gautier}, {Gomillion}, \&
  {Wilkins}}]{2011ApJ...741...68M}
{Masiero}, J.~R., {Mainzer}, A.~K., {Grav}, T., {et~al.} 2011, \apj, 741, 68,
  \dodoi{10.1088/0004-637X/741/2/68}

\bibitem[{{Miljkovi{\'c}} {et~al.}(2016){Miljkovi{\'c}}, {Collins},
  {Wieczorek}, {Johnson}, {Soderblom}, {Neumann}, \&
  {Zuber}}]{2016JGRE..121.1695M}
{Miljkovi{\'c}}, K., {Collins}, G.~S., {Wieczorek}, M.~A., {et~al.} 2016,
  Journal of Geophysical Research (Planets), 121, 1695,
  \dodoi{10.1002/2016JE005038}

\bibitem[{{Minton} {et~al.}(2019{\natexlab{a}}){Minton}, {Fassett},
  {Hirabayashi}, \& {Riedel}}]{2019EPSC...13.1065M}
{Minton}, D., {Fassett}, C., {Hirabayashi}, M., \& {Riedel}, C.
  2019{\natexlab{a}}, in EPSC-DPS Joint Meeting 2019, Vol. 2019,
  EPSC--DPS2019--1065

\bibitem[{{Minton} {et~al.}(2019{\natexlab{b}}){Minton}, {Fassett},
  {Hirabayashi}, {Howl}, \& {Richardson}}]{2019Icar..326...63M}
{Minton}, D.~A., {Fassett}, C.~I., {Hirabayashi}, M., {Howl}, B.~A., \&
  {Richardson}, J.~E. 2019{\natexlab{b}}, \icarus, 326, 63,
  \dodoi{10.1016/j.icarus.2019.02.021}

\bibitem[{{Minton} \& {Malhotra}(2010)}]{2010Icar..207..744M}
{Minton}, D.~A., \& {Malhotra}, R. 2010, \icarus, 207, 744,
  \dodoi{10.1016/j.icarus.2009.12.008}

\bibitem[{{Morbidelli} {et~al.}(2010){Morbidelli}, {Brasser}, {Gomes},
  {Levison}, \& {Tsiganis}}]{2010AJ....140.1391M}
{Morbidelli}, A., {Brasser}, R., {Gomes}, R., {Levison}, H.~F., \& {Tsiganis},
  K. 2010, \aj, 140, 1391, \dodoi{10.1088/0004-6256/140/5/1391}

\bibitem[{{Morbidelli} {et~al.}(2009){Morbidelli}, {Brasser}, {Tsiganis},
  {Gomes}, \& {Levison}}]{2009A&A...507.1041M}
{Morbidelli}, A., {Brasser}, R., {Tsiganis}, K., {Gomes}, R., \& {Levison},
  H.~F. 2009, \aap, 507, 1041, \dodoi{10.1051/0004-6361/200912876}

\bibitem[{{Morbidelli} {et~al.}(2015){Morbidelli}, {Walsh}, {O'Brien},
  {Minton}, \& {Bottke}}]{2015aste.book..493M}
{Morbidelli}, A., {Walsh}, K.~J., {O'Brien}, D.~P., {Minton}, D.~A., \&
  {Bottke}, W.~F. 2015, in Asteroids IV, ed. P.~{Michel}, F.~E. {DeMeo}, \&
  W.~F. {Bottke}, 493--507, \dodoi{10.2458/azu_uapress_9780816532131-ch026}

\bibitem[{{Nesvorn{\'y}}(2015)}]{2015PDSS..234.....N}
{Nesvorn{\'y}}, D. 2015, NASA Planetary Data System, EAR

\bibitem[{{Nesvorn{\'y}}(2018)}]{2018ARA&A..56..137N}
---. 2018, \araa, 56, 137, \dodoi{10.1146/annurev-astro-081817-052028}

\bibitem[{{Nesvorn{\'y}} {et~al.}(2015){Nesvorn{\'y}}, {Bro{\v{z}}}, \&
  {Carruba}}]{2015aste.book..297N}
{Nesvorn{\'y}}, D., {Bro{\v{z}}}, M., \& {Carruba}, V. 2015, in Asteroids IV,
  ed. P.~{Michel}, F.~E. {DeMeo}, \& W.~F. {Bottke}, 297--321,
  \dodoi{10.2458/azu_uapress_9780816532131-ch016}

\bibitem[{{Nesvorn{\'y}} \& {Morbidelli}(2012)}]{2012AJ....144..117N}
{Nesvorn{\'y}}, D., \& {Morbidelli}, A. 2012, \aj, 144, 117,
  \dodoi{10.1088/0004-6256/144/4/117}

\bibitem[{{Nesvorn{\'y}} \& {Roig}(2018)}]{2018AJ....155...42N}
{Nesvorn{\'y}}, D., \& {Roig}, F. 2018, \aj, 155, 42,
  \dodoi{10.3847/1538-3881/aa9a47}

\bibitem[{{Nesvorn{\'y}} {et~al.}(2017){Nesvorn{\'y}}, {Roig}, \&
  {Bottke}}]{2017AJ....153..103N}
{Nesvorn{\'y}}, D., {Roig}, F., \& {Bottke}, W.~F. 2017, \aj, 153, 103,
  \dodoi{10.3847/1538-3881/153/3/103}

\bibitem[{{Nesvorn{\'y}} {et~al.}(2008){Nesvorn{\'y}}, {Roig}, {Gladman},
  {Lazzaro}, {Carruba}, \& {Moth{\'e}-Diniz}}]{2008Icar..193...85N}
{Nesvorn{\'y}}, D., {Roig}, F., {Gladman}, B., {et~al.} 2008, \icarus, 193, 85,
  \dodoi{10.1016/j.icarus.2007.08.034}

\bibitem[{{Nesvorn{\'y}} {et~al.}(2018){Nesvorn{\'y}}, {Vokrouhlick{\'y}},
  {Bottke}, \& {Levison}}]{2018NatAs...2..878N}
{Nesvorn{\'y}}, D., {Vokrouhlick{\'y}}, D., {Bottke}, W.~F., \& {Levison},
  H.~F. 2018, Nature Astronomy, 2, 878, \dodoi{10.1038/s41550-018-0564-3}

\bibitem[{{Neukum} {et~al.}(2001){Neukum}, {Ivanov}, \&
  {Hartmann}}]{2001SSRv...96...55N}
{Neukum}, G., {Ivanov}, B.~A., \& {Hartmann}, W.~K. 2001, \ssr, 96, 55

\bibitem[{{O'Brien} {et~al.}(2006){O'Brien}, {Greenberg}, \&
  {Richardson}}]{2006Icar..183...79O}
{O'Brien}, D.~P., {Greenberg}, R., \& {Richardson}, J.~E. 2006, \icarus, 183,
  79, \dodoi{10.1016/j.icarus.2006.02.008}

\bibitem[{{O'Brien} {et~al.}(2014){O'Brien}, {Marchi}, {Morbidelli}, {Bottke},
  {Schenk}, {Russell}, \& {Raymond}}]{2014P&SS..103..131O}
{O'Brien}, D.~P., {Marchi}, S., {Morbidelli}, A., {et~al.} 2014, \planss, 103,
  131, \dodoi{10.1016/j.pss.2014.05.013}

\bibitem[{{O'Brien} {et~al.}(2007){O'Brien}, {Morbidelli}, \&
  {Bottke}}]{2007Icar..191..434O}
{O'Brien}, D.~P., {Morbidelli}, A., \& {Bottke}, W.~F. 2007, \icarus, 191, 434,
  \dodoi{10.1016/j.icarus.2007.05.005}

\bibitem[{{Petit} {et~al.}(2001){Petit}, {Morbidelli}, \&
  {Chambers}}]{2001Icar..153..338P}
{Petit}, J.-M., {Morbidelli}, A., \& {Chambers}, J. 2001, \icarus, 153, 338,
  \dodoi{10.1006/icar.2001.6702}

\bibitem[{{Robbins}(2014)}]{2014E&PSL.403..188R}
{Robbins}, S.~J. 2014, Earth and Planetary Science Letters, 403, 188,
  \dodoi{10.1016/j.epsl.2014.06.038}

\bibitem[{{Roig} \& {Nesvorn{\'y}}(2015)}]{2015AJ....150..186R}
{Roig}, F., \& {Nesvorn{\'y}}, D. 2015, \aj, 150, 186,
  \dodoi{10.1088/0004-6256/150/6/186}

\bibitem[{{Scheeres} {et~al.}(2015){Scheeres}, {Britt}, {Carry}, \&
  {Holsapple}}]{2015aste.book..745S}
{Scheeres}, D.~J., {Britt}, D., {Carry}, B., \& {Holsapple}, K.~A. 2015, in
  Asteroids IV, ed. P.~{Michel}, F.~E. {DeMeo}, \& W.~F. {Bottke}, 745--766,
  \dodoi{10.2458/azu_uapress_9780816532131-ch038}

\bibitem[{{Schmedemann} {et~al.}(2014){Schmedemann}, {Kneissl}, {Ivanov},
  {Michael}, {Wagner}, {Neukum}, {Ruesch}, {Hiesinger}, {Krohn}, {Roatsch},
  {Preusker}, {Sierks}, {Jaumann}, {Reddy}, {Nathues}, {Walter}, {Neesemann},
  {Raymond}, \& {Russell}}]{2014P&SS..103..104S}
{Schmedemann}, N., {Kneissl}, T., {Ivanov}, B.~A., {et~al.} 2014, \planss, 103,
  104, \dodoi{10.1016/j.pss.2014.04.004}

\bibitem[{{Shoemaker}(1962)}]{1962pam..book.....K}
{Shoemaker}, E.~M. 1962, in Physics and astronomy of the moon, ed. Z.~{Kopal}
  (New York: Academic Press), 283--359

\bibitem[{{Thomas} {et~al.}(1997){Thomas}, {Binzel}, {Gaffey}, {Zellner},
  {Storrs}, \& {Wells}}]{1997Icar..128...88T}
{Thomas}, P.~C., {Binzel}, R.~P., {Gaffey}, M.~J., {et~al.} 1997, \icarus, 128,
  88, \dodoi{10.1006/icar.1997.5736}

\bibitem[{{Vokrouhlick{\'y}} {et~al.}(2015){Vokrouhlick{\'y}}, {Bottke},
  {Chesley}, {Scheeres}, \& {Statler}}]{2015aste.book..509V}
{Vokrouhlick{\'y}}, D., {Bottke}, W.~F., {Chesley}, S.~R., {Scheeres}, D.~J.,
  \& {Statler}, T.~S. 2015, in Asteroids IV, ed. P.~{Michel}, F.~E. {DeMeo}, \&
  W.~F. {Bottke}, 509--531, \dodoi{10.2458/azu_uapress_9780816532131-ch027}

\bibitem[{{Vokrouhlick{\'y}} {et~al.}(2006){Vokrouhlick{\'y}}, {Bro{\v{z}}},
  {Morbidelli}, {Bottke}, {Nesvorn{\'y}}, {Lazzaro}, \&
  {Rivkin}}]{2006Icar..182...92V}
{Vokrouhlick{\'y}}, D., {Bro{\v{z}}}, M., {Morbidelli}, A., {et~al.} 2006,
  \icarus, 182, 92, \dodoi{10.1016/j.icarus.2005.12.011}

\bibitem[{{Walsh} {et~al.}(2011){Walsh}, {Morbidelli}, {Raymond}, {O'Brien}, \&
  {Mandell}}]{2011Natur.475..206W}
{Walsh}, K.~J., {Morbidelli}, A., {Raymond}, S.~N., {O'Brien}, D.~P., \&
  {Mandell}, A.~M. 2011, \nat, 475, 206, \dodoi{10.1038/nature10201}

\bibitem[{{Weidenschilling}(1977)}]{1977Ap&SS..51..153W}
{Weidenschilling}, S.~J. 1977, \apss, 51, 153, \dodoi{10.1007/BF00642464}

\end{thebibliography}

\begin{table}
\caption{Values of different parameters adopted in the present study. See \citet{2016Sci...353.4759H} for 
$D_{\mathrm{sc}}$ and \citet{2015aste.book..745S} for $\delta$.}
\label{params}
\centering{}%
\begin{tabular}{lcc}
\hline 
 & Ceres & Vesta\tabularnewline
\hline 
Impactor density $\delta$ (g~cm$^{-3}$) & \multicolumn{2}{c}{2.5}\tabularnewline
Target density $\rho$ (g~cm$^{-3}$) & 2.08 & 3.42\tabularnewline
Surface gravity $g$ ($\mathrm{m\,s}{}^{-2}$) & 0.273 & 0.253\tabularnewline
Simple to complex transition $D_{\mathrm{sc}}$ (km) & 10 & 60\tabularnewline
Target surface $S$ ($\times10^{5}$ km$^{2}$) & 27.7 & 8.8\tabularnewline
Minimum impactor diameter $D_{\mathrm{ast},0}$ (km) & 4 & 6\tabularnewline\hline 
\end{tabular}
\end{table}

\begin{table}
\caption{A comparison between the number of observed and model predicted craters for Vesta and Ceres.
Each predicted number represents the expected value of a Poisson distribution over 4.56 Gy. The values for the early and late instability cases are reported. Here we do not account 
for any crater erasure process (Sect. \ref{subsec:mpf})}.
\label{tab-counts}
\centering{}%
\begin{tabular}{ccccccccc}
\hline 
 &  & \multicolumn{3}{c}{Vesta} &  & \multicolumn{3}{c}{Ceres}\tabularnewline
\cline{3-5} \cline{7-9} 
Impactors &  & Craters & \multirow{2}{*}{Observed$^{\dagger}$} & Predicted &  & Craters & \multirow{2}{*}{Observed$^{\ddagger}$} & Predicted\tabularnewline
(km) &  & (km) &  & Early / Late &  & (km) &  & Early / Late\tabularnewline
\hline 
$D_{\mathrm{ast}}>8$ &  & $D_{\mathrm{crat}}>70$ & - & 22 / 28 &  & $D_{\mathrm{crat}}>90$ & 26 & 68 / 69\tabularnewline
$D_{\mathrm{ast}}>9$ &  & $D_{\mathrm{crat}}>80$ & - & 16 / 21 &  & $D_{\mathrm{crat}}>100$ & 17 & 51 / 52\tabularnewline
$D_{\mathrm{ast}}>10$ &  & $D_{\mathrm{crat}}>90$ & 10 & 13 / 16 &  & $D_{\mathrm{crat}}>110$ & 14 & 40 / 41\tabularnewline
$D_{\mathrm{ast}}>11$ &  & $D_{\mathrm{crat}}>100$ & - & 10 / 13 &  & $D_{\mathrm{crat}}>120$ & 8 & 32 / 33\tabularnewline
$D_{\mathrm{ast}}>12$ &  & $D_{\mathrm{crat}}>110$ & 6 & 8 / 11 &  & $D_{\mathrm{crat}}>130$ & 6 & 26 / 27\tabularnewline
\hline 
\end{tabular}
\centering{}$^{\dagger}$\citet{2012Sci...336..690M}\\
\centering{}$^{\ddagger}$\citet{2016NatCo...712257M}
\end{table}

\begin{table}
	\caption{The probability of forming basins with $D_{\mathrm{crat}}>400$ km, obtained from our model production function for Vesta using the Poisson statistics. The early instability is assumed here, and no crater erasure is accounted for.}
	\label{tab-poison}
	\centering{}%
	\begin{tabular}{lccc}
		\hline 
		Isochrone (Ga) & $\mu$ & $p_{\mu}(1)$ & $p_{\mu}(2)$ \tabularnewline
		\hline 
		$T< 1.0$ & $0.16$ & $13$\% & $1$\%\tabularnewline
        $T< 2.0$ & $0.28$ & $21$\% & $3$\%\tabularnewline
        $T< 3.0$ & $0.40$ & $27$\% & $5$\%\tabularnewline
		$T< 4.56$ & $0.65$ & $34$\% & $11$\%\tabularnewline
		\hline 
	\end{tabular}
\end{table}

\begin{table}
	\caption{The probability of forming Rheasilvia and Venenia basins ($D_{\mathrm{crat}}>400$ km), assuming that the former formed at $T \leq 1$ Ga and considering different age intervals for the later.}
	\label{tab-bene}
	\centering{}%
	\begin{tabular}{ccc}
		\hline 
		Venenia Age (Ga) & Early Instability & Late Instability\tabularnewline
		\hline 
		$1.0<T<4.56$ & $6.0$\% & $4.0$\%\tabularnewline
		$1.0<T<3.0$ & $2.0$\% & $1.0$\%\tabularnewline
		$1.0<T<2.0$ & $1.5$\% & $0.0$\%\tabularnewline
		\hline 
	\end{tabular}
\end{table}

\end{document}